\documentclass[dvipsnames,twocolumn]{aastex631}

\usepackage{graphicx}
\usepackage{amsmath}
\usepackage{txfonts}
\usepackage{color}
\usepackage{comment}
\usepackage{mathtools}  
\usepackage{xfrac}

\graphicspath{{Figures/}}

\newcommand{\newtext}[1]{#1}

\newcommand{\nhp}[0]{N$_2$H$^+$ }
\newcommand{\cyo}[0]{C$^{18}$O}
\newcommand{\xco}[0]{$^{13}$CO}
\newcommand{\co}[0]{$^{12}$CO}

\newcommand{\mdisk}[0]{\ensuremath{M_{\rm disk}}}
\newcommand{\mgas}[0]{\ensuremath{M_{\rm gas}}}
\newcommand{\mdust}[0]{\ensuremath{M_{\rm dust}}}
\newcommand{\rgas}[0]{\ensuremath{R_{\rm CO,\ 90\%}}}
\newcommand{\rc}[0]{\ensuremath{R_{\rm c}}}

\newcommand{\msun}[0]{\ensuremath{\mathrm{M}_{\odot}}}

\newcommand{\zfreeze}[0]{\ensuremath{z_{\rm freeze}}}

\begin{document} 

    \title{How large is a disk - what do protoplanetary disk gas sizes really mean?}

\correspondingauthor{Leon Trapman}
\email{ltrapman@wisc.edu}
\author[0000-0002-8623-9703]{Leon Trapman}
\affiliation{Department of Astronomy, University of Wisconsin-Madison, 
475 N Charter St, Madison, WI 53706}

\author[0000-0003-4853-5736]{Giovanni Rosotti}
\affiliation{Dipartimento di Fisica, Universit\`a degli Studi di Milano, Via Giovanni Celoria, 16, 20133, Milano, Italy}
\affiliation{School of Physics and Astronomy, University of Leicester, Leicester LE1 7RH, UK}
\affiliation{Leiden Observatory, Leiden University, 2300 RA Leiden, the Netherlands}

\author[0000-0002-0661-7517]{Ke Zhang}
\affiliation{Department of Astronomy, University of Wisconsin-Madison, 
475 N Charter St, Madison, WI 53706}

\author[0000-0002-1103-3225]{Beno\^it Tabone}
\affiliation{Universit\'e Paris-Saclay, CNRS, Institut d’Astrophysique Spatiale, F-91405 Orsay, France}
\affiliation{Leiden Observatory, Leiden University, 2300 RA Leiden, the Netherlands}

\begin{abstract}
It remains unclear what mechanism is driving the evolution of protoplanetary disks. Direct detection of the main candidates, either turbulence driven by magnetorotational instability or magnetohydrodynamical disk winds, has proven difficult, leaving the time evolution of the disk size as one of the most promising observables able to differentiate between these two mechanisms. But to do so successfully, we need to understand what the observed gas disk size actually traces. We studied the relation between $R_{\rm CO,\ 90\%}$, the radius that encloses 90\% of the $^{12}$CO flux, and $R_c$, the radius that encodes the physical disk size, in order to provide simple prescriptions for conversions between these two sizes. For an extensive grid of thermochemical models we calculate $R_{\rm CO,\ 90\%}$ from synthetic observations and relate properties measured at this radius, such as the gas column density, to bulk disk properties, such as $R_c$ and the disk mass $M_{\rm disk}$. We found an empirical correlation between the gas column density at $R_{\rm CO,\ 90\%}$ and disk mass: $N_{\rm gas}(R_{\rm CO,\ 90\%}) \approx 3.73\times10^{21}(M_{\rm disk}/\mathrm{M}_{\odot})^{0.34}\ \mathrm{cm}^{-2}$. Using this correlation we derive an analytical prescription of $R_{\rm CO,\ 90\%}$ that only depends on $R_c$ and $M_{\rm disk}$. 
We derive $R_c$ for disks in Lupus, Upper Sco, Taurus and DSHARP, finding that disks in the older Upper Sco region are significantly smaller ($\langle R_c \rangle$ = 4.8 au) than disks in the younger Lupus and Taurus regions ($\langle R_c \rangle$ = 19.8 and 20.9 au, respectively). This temporal decrease in $R_c$ goes against predictions of both viscous and wind-driven evolution, but could be a sign of significant external photoevaporation having truncated disks in Upper Sco.
\end{abstract}

\section{Introduction}
\label{sec: introduction}

Proto-planetary disks are the birth-sites of planets and only by understanding disks and their properties can we understand planet formation \citep[e.g.,][]{MorbidelliRaymond2016}.

Among the disk properties, size is one of the most fundamental. On a simple level, in combination with the disk mass, disk size is the main parameter determining the disk surface density, which in turn represent the available material to be accreted into planets. On a perhaps deeper level, the evolution of the size can inform us on the mechanism driving disc evolution. For example, in a scenario in which accretion is driven by viscosity, the disc size needs to get larger with time \citep{LyndenBellPringle1974,Hartmann1998} in order to conserve the disk angular momentum: this is normally called viscous spreading. Conversely, if angular momentum is extracted by MHD winds, expansion is not required (\citealt{Armitage2013,Bai2016,Tabone2022a}\newtext{; although see \citealt{YangBai2021} for the possibility of wind-driven disks growing over time}). \newtext{It is worth mentioning that both processes could affect different parts of the disk simultaneously, thereby complicating our simple view of disk evolution (e.g. \citealt{AlessiPudritz2022})} Other mechanisms such as the presence of a stellar companion \citep[e.g.][]{Papaloizou1977,ArtymowiczLubow1994,RosottiClarke2018,Zagaria2021,ZagariaReview}, external photo-evaporation \citep[e.g.,][]{Clarke2007,Facchini2016,Haworth2018,Sellek2020,WinterReview} and, if the disk size is determined from the dust \newtext{continuum emission at sub-millimeter wavelengths}, radial drift \citep{Weidenschilling1977,Rosotti2019} can reduce the size of a proto-planetary disk and make it smaller with time, which has important consequences for disk evolution.

In the previous discussion we have been purposely negligent in describing in detail what ``size'' means. The underlying assumption in the way the term is normally used is that the disk size should somehow reflect where the disk mass is distributed. In practice, since following the analytical solutions of \citet{LyndenBellPringle1974} it is common to parameterize disk surface densities using an exponentially tapered power-law, disk size is often intended as the scale radius of the exponential, normally denoted with  $R_{\rm c}$. Any other parametrization of the surface density can always be characterized by defining the radius enclosing a given fraction of the disk mass.

In observations, however, proto-planetary disks have multiple ``sizes'', and one has to be careful which size is being considered for any analysis to be meaningful. Sizes are different first of all because proto-planetary disks can be observed at multiple wavelengths and in multiple tracers. Before ALMA became available, most available measurements of disk sizes were done in the continuum at sub-mm wavelengths (see review of pre-ALMA results by \citealt{WilliamsCieza2011}), with measurements available also at optical wavelength thanks to HST \citep{VincenteAlves2005}, although predominantly for objects in Orion. While ALMA greatly expanded the sample of sub-mm continuum disc sizes \citep[e.g.,][]{Andrews2018,Hendler2020,ManaraPPVII,Tazzari2021}), one of ALMA biggest contributions is that we now have relatively large samples with measurements of gas sizes \citep{ansdell2017,barenfeld2017,Sanchis2021,Long2022}. We should highlight however that also ``gas'' is a generic term since many different gas-phase species are known in proto-planetary disks. In this paper with ``gas'' disk size we always mean its most abundant species, CO, and particularly its most abundant isotopologue, $^{12}$CO. This choice is motivated by the fact by far $^{12}$CO, in virtue of its brightness, is the tracer with the largest observational sample of measured disk sizes.

Even once the wavelength and tracer are specified, one still needs to specify how the disk size is exactly determined from the observations - e.g., see \citet{Tripathi2017} for a discussion concerning the continuum. In this paper we will consider as observational disk size the radius enclosing a given fraction of the total flux, since this definition is generic enough to be applied to any observation, and following common observational conventions take the fraction to be 90 \%. We denote this radius as \rgas.

Regardless of the observational tracer, one should stress that no available tracer is really tracing the disk size in the purely theoretical sense; i.e., these tracers tell us the surface brightness distribution of the given tracer, and not how the mass of the disk is distributed. 
This is because of several reasons: the abundance of the chosen tracer may vary throughout, the intensity can get weaker or stronger as the disk temperature varies, and the given tracer may not be optically thin, implying that its surface brightness does not trace its surface density.
Investigating the link between the observed size (\rgas) of a proto-planetary disk and the theoretical size (\rc) is the purpose of this paper.

In order to accomplish this goal, we have run a grid of thermochemical models where we compute the abundance of $^{12}$CO in the disk and we have ray-traced the models to account for radiative transfer effects. Starting from earlier work presented in \citet{Trapman2022} and \citet{Toci2023}, we then use this grid to derive simple, yet accurate, analytical relations which allow us to predict the observed disk size for a given disk mass and theoretical size. The benefit of an analytical relation is that it can be inverted relatively easily. We make use of this to derive \rc\ from observations of \rgas\ of disks in Lupus and Upper Sco, and discuss the implications for disk evolution.

The paper is structured as follows. We first present the technical details of our models in section \ref{sec: models}, and then show our results concerning the relation between \rc\ and \rgas\ in section \ref{sec: results}. In section \ref{sec: discussion} we apply the inverse relation to measure \rc\ in an observational sample and discuss the caveats of our work, before finally drawing our conclusions in section \ref{sec: conclusions}.

\section{The \texttt{DALI} models}
\label{sec: models}

The location of \rgas, defined as the radius that encloses 90\% of the $^{12}$CO 2-1 flux, depends on the CO emission profile, which in turn depends on the CO chemistry and thermal structure of the disk, both of which can be obtained using a thermochemical model. In this work we use the thermochemical code \texttt{DALI} \citep{Bruderer2012,Bruderer2013} to run a series of disk models.
\texttt{DALI} self-consistently calculates the thermal and chemical structure of a disk with a given (gas and dust) density structure and stellar radiation field. The code first computes the internal radiation field and dust temperature structure using a 2D Monte Carlo method to solve the radiative transfer equation. It then iteratively solves the time-dependent chemistry, calculates molecular and atomic excitation levels, and computes the gas temperature by balancing heating and cooling processes until a self-consistent solution is found. Finally, the model is raytraced to construct synthetic emission maps. A more detailed description of the code is provided in Appendix A of \cite{Bruderer2012}.

For the surface density profile of our models we take the self-similar solution of the generalized, i.e. viscous and/or wind-driven, disk evolution given in \cite{Tabone2022a}, which is a tapered power-law of the form 
\begin{equation}
\label{eq: surface density}
\Sigma_{\rm gas}(R) = \Gamma\left(\frac{\xi + 2 -\gamma}{2-\gamma}\right) \frac{\mdisk}{2\pi\rc^2} \left( \frac{R}{\rc}\right)^{-\gamma+\xi} \exp \left[-\left(\frac{R}{\rc}\right)^{2-\gamma}\right].
\end{equation}
Here $\mdisk$ is the mass of the disk, $\rc$ is the characteristic size, $\gamma$ \newtext{is the slope of the surface density, which} is related to the slope of $\tilde{\alpha}$ (see \citealt{Tabone2022a}). \newtext{For the viscous case $\gamma$} coincides with the slope of the kinetic viscosity \newtext{(see, e.g. \citealt{LyndenBellPringle1974})}.
$\xi$ is the mass ejection index \citep{FerreiraPelletier1995,Ferreira1997} and $\Gamma$ is the gamma function, which for common ranges of $\gamma$ and $\xi$ is a factor of order unity.
In this work we will set $\xi=0.25$, which is equivalent with only vertical angular momentum transport by a MHD wind. 
Note that $\xi$ has only a small effect on \rgas\ as shown in Figure 11 in \cite{Trapman2022}. Similarly we set $\gamma=1$ for most of this work, but see in Section \ref{sec: correlation -- dependencies} for the effect of $\gamma$ on our results. \newtext{Note that in contrast to \cite{Trapman2020,Trapman2022} disk evolution is not included and the surface density is fixed for each model.}

The vertical density is assumed to be a Gaussian around disk midplane, which is the outcome of hydrostatic equilibrium under the simplifying assumption that the disk is vertically isothermal (see Eq \eqref{eq: vertical gaussian}). To simulate the effect of observed disk flaring (e.g., \citealt{DullemondDominik2004,avenhaus2018,Law2021bMAPS,Law2022}), the vertical scale height of the disk is described by a powerlaw
\begin{equation}
\label{eq: scale height}
H(R) = R h_c \left(\frac{R}{R_c}\right)^{\psi}
\end{equation}
where $h_c$ is the opening angle at \rc\ and $\psi$ is the flaring angle.

Dust is included in the form of two dust population following e.g. \cite{Andrews2011}. Small grains [0.005-1 $\mu$m], making up a fraction $(1-f_{\rm large})$ of the total dust mass are distributed over the full vertical and radial extent of the disk, following the gas. Large grains [1-$10^3\mu$m] that make up the remaining $f_{\rm large}$ fraction of the dust mass have the same radial distribution as the gas, but are vertically confined to the midplane to simulate the effect of vertical dust settling. This is achieved by reducing their scale height by a factor $\chi < 1$. 

Finally, the star is assumed to be a 4000 K blackbody with a stellar radius chosen such that the star has a stellar luminosity $L_* = 0.28\ \mathrm{L}_{\odot}$. To this spectrum we add a $10^4$ K blackbody to simulate the accretion luminosity released by a $10^{-8}\ \msun/\mathrm{yr}$ stellar mass accretion flow, where we assume that 50\% of the gravitational potential energy is released as radiation (e.g. \citealt{kama2015}).  
Table \ref{tab: model fixed parameters} summarizes the parameters of our fiducial models. 

To test the empirical correlation presented in the next section we also ran multiple sets of models that similar to our fiducial models span a range of disk masses but where one of the fiducial model parameters was varied over two or more values. \newtext{The selected parameters are all expected to have a significant effect on the gas density, the temperature structure and/or the chemistry of CO.} These model parameters include: the stellar luminosity $L_*$, the opening angle $h_c$, the external UV field (ISRF), the characteristic radius \rc, the slope of the surface density $\gamma$, the flaring angle $\psi$, the dust settling parameter $\chi$ and the fraction of large grains $f_{\rm large}$. 
\newtext{Further parameters such as, for example, the UV luminosity of the star were also examined, but tests showed that they had no significant effect on \rgas.}
\newtext{The inclination of the disk can also affect \rgas, but its effects can be minimized for moderately inclined disks ($< 60$ deg) by measuring \rgas\ in the deprojected disk frame (see, e.g. appendix A in \citealt{Trapman2019})}.

\begin{table}[tbh]
  \centering   
  \caption{\label{tab: model fixed parameters}Fiducial \texttt{DALI} model parameters.}
  \begin{tabular*}{0.8\columnwidth}{ll}
    \hline\hline
    Parameter & Range\\
    \hline
     \textit{Chemistry}$^{1}$&\\
     Chemical age & 1 Myr\\
     {[C]/[H]}$^{1}$ & $1.35\cdot10^{-4}$\\
     {[O]/[H]} & $2.88\cdot10^{-4}$\\
     \textit{Physical structure} &\\ 
     $\gamma$ &  [0.5, \textbf{1.0}, 1.5]\\
     $\xi$ & 0.25\\
     $\psi$ & [0.05, \textbf{0.15}, 0.25]\\ 
     $h_c$ &  [\textbf{0.1}, 0.2] \\ 
     $R_c$ & [5, 20, 40, \textbf{65}] au\\
     $M_{\mathrm{gas}}$ & $5\times10^{-7} - 10^{-1}$ M$_{\odot}$ \\
     Gas-to-dust ratio & 100 \\
     \textit{Dust properties} & \\
     $f_{\mathrm{large}}$ & [0.8, \textbf{0.9}, 0.99] \\
     $\chi$ & [0.1, \textbf{0.2}, 0.4] \\
     composition & standard ISM$^{2}$\\
     \textit{Stellar spectrum} & \\
     $T_{\rm eff}$ & 4000 K + Accretion UV \\
     $L_{*}$ & [0.1, \textbf{0.28}, 1.0, 3.0] L$_{\odot}$  \\
     $\zeta_{\rm cr}$ & $10^{-17}\ \mathrm{s}^{-1}$\\
     \textit{Observational geometry}&\\
     $i$ & 0$^{\circ}$ \\
     PA & 0$^{\circ}$ \\
     $d$ & 150 pc\\
    \hline
  \end{tabular*}
  \begin{minipage}{0.75\columnwidth}
  \vspace{0.1cm}
  {\footnotesize{$^1$\newtext{We assume typical ISM abundances for the total carbon and oxygen abundances \citep{Cardelli1996,Jonkheid2007,Woitke2009,Bruderer2012}} $^{2}$\citealt{WeingartnerDraine2001}, see also Section 2.5 in \citealt{Facchini2017}. Parameters shown in bold are varied in Section \ref{sec: correlation -- dependencies}. }}
  \end{minipage}
\end{table}

\section{Results}
\label{sec: results}

\begin{figure*}[htb]
    \centering
    \includegraphics[width=\textwidth]{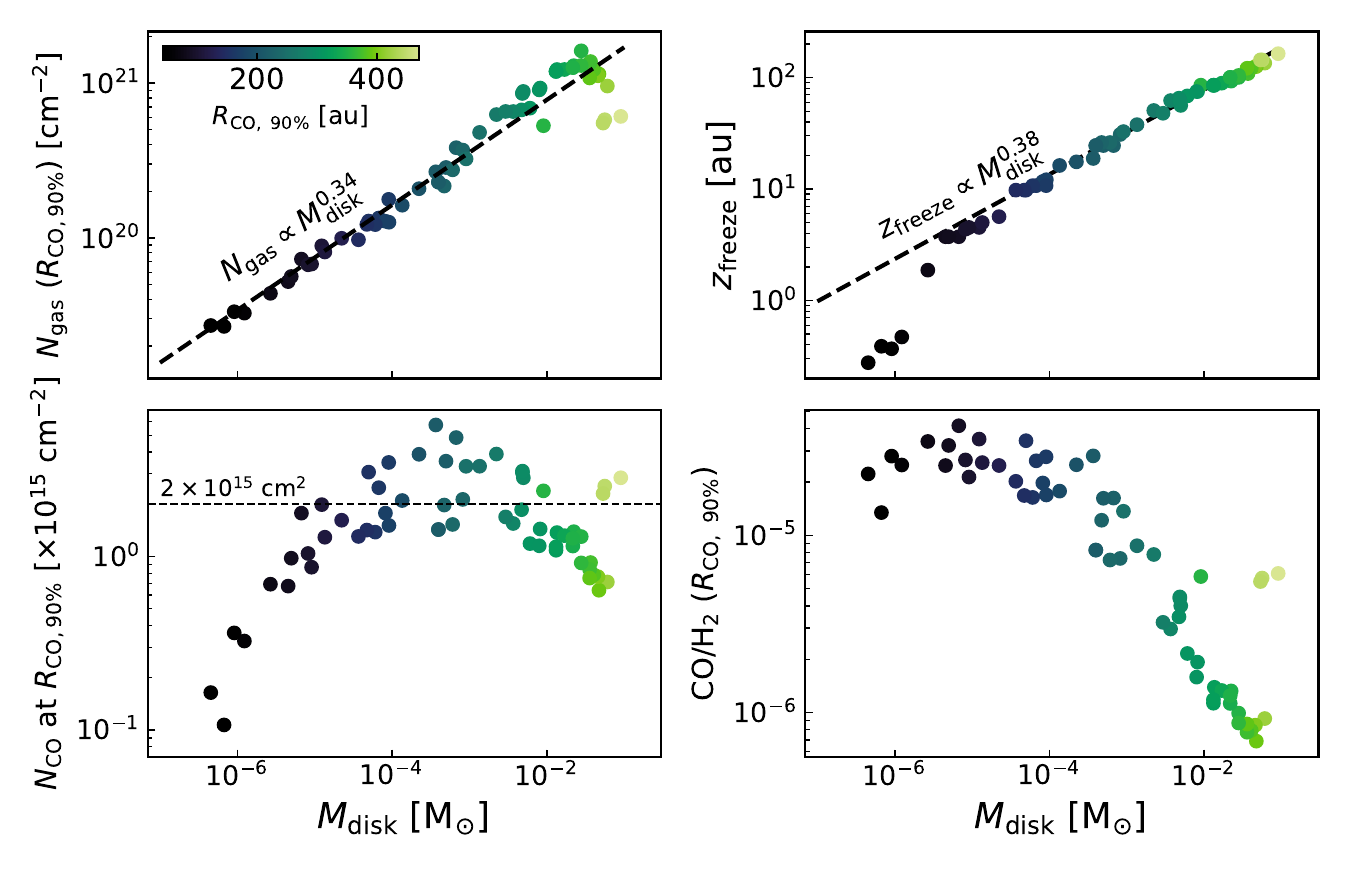}
    \caption{\label{fig: Ngas-mdisk correlation}Properties measured at \rgas\ from DALI models with $\rc=65$ au and $\mdisk = [10^{-7}-10^{-1}\ \mathrm{M}_{\odot}]$. \textbf{Top left:} Gas column density at \rgas\ against the disk mass. Colors show the \rgas\ of the disk. The black dashed line shows the correlation between $N_{\rm gas}(\rgas)$ and \mdisk. \textbf{Top right:} The height of the CO freeze-out layer at \rgas\ versus the disk mass. The black dashed line shows a correlation between \zfreeze(\rgas) and \mdisk, similar to the one seen in the top left panel. \textbf{Bottom left:} CO column density at \rgas\ versus the disk mass. \textbf{Bottom right:} Column-averaged CO abundance (i.e. $N_{\rm CO}(\rgas)/N_{\rm gas}(\rgas)$ at \rgas\ versus disk mass. }
\end{figure*}

\subsection{A tight empirical correlation between $N_{\rm gas}(\rgas)$ and the disk mass}
\label{sec: correlation -- first look}

It is common practice to measure the protoplanetary gas disk sizes from the extent of the sub-millimeter $^{12}$CO rotational emission. These low $J$ lines require a relatively small column to become optically thick, allowing us to easily detect the low density material found in the outer part of the disk. Furthermore, at low column densities UV photons are able to photo-dissociate CO, thus removing the molecule from the gas. The exact CO column density required to self-shield against this depends somewhat on the molecular hydrogen column and the temperature, but it lies at around a few times $10^{15}\ \mathrm{cm}^{-2}$ (see, e.g. \citealt{vanDishoeckBlack1988}). Back-of-the-envelope calculations show that the radius where CO millimeter lines becomes optically thin $(R_{\tau[mm]=1})$ approximately coincides with the radius where it stops being able to self-shield against photodissociation $(R_{\rm CO\ p.d.})$. 
It should be noted that CO is also partly protected by mutual line shielding of CO by H$_2$, but this is negligible compared to the effect of CO self-shielding (see, e.g. \citealt{Lee1996}). 
This sets the expectation of a link between the observed gas disk size \rgas, which is linked to $R_{\tau=1}$, and the surface density, albeit indirectly, from $N_{\rm CO}(R_{\rm CO\ p.d.})\approx 10^{15}\ \mathrm{cm}^{-2}$ (see, e.g., \citealt{Toci2023,Trapman2022}).

Using our thermochemical models, we can test this expectation. After measuring \rgas\ from the synthetic CO 2-1 observations of our models we find a surprisingly tight correlation between the gas column density\footnote{In this work the gas column density is defined assuming a mean molecular weight $\mu=2.3$, so $N_{\rm gas} = \tfrac{\Sigma_{\rm gas}}{2.3 m_H}$.} at the observed outer radius $(N_{\rm gas}(\rgas))$ and the mass of the disk (\mdisk). The top left panel of Figure \ref{fig: Ngas-mdisk correlation} shows that $N_{\rm gas}(\rgas)$ increases with \mdisk\ as a powerlaw, $N_{\rm gas}(\rgas) \propto \mdisk^{0.34}$. 

The positive correlation can be understood, at least qualitatively, by looking at the other quantities shown in Figure \ref{fig: Ngas-mdisk correlation} that are also obtained at \rgas. 
First off, the column density of CO at \rgas, denoted as $N_{\rm CO}(\rgas)$, has an approximately constant value of $\approx 2\times10^{15}\ \mathrm{cm}^{-2}$ across the full disk mass range examined here.
This value corresponds to the CO column required for CO self-shielding (e.g. \citealt{vanDishoeckBlack1988}), which matches with the expectation discussed earlier that \rgas\ roughly coincides the radius where CO starts to become photo-dissociated. We would therefore expect that the observed disk size \rgas increases with disk mass, because this critical CO column density, assuming a fixed CO abundance, lies further outward for a disk that has more mass (see e.g. \citealt{Trapman2020}). However, further out from the star the disk is also colder and a larger fraction of the CO column is frozen out, resulting in a lower column-averaged CO abundance. This is corroborated by the rightmost panels of Figure \ref{fig: Ngas-mdisk correlation}, which show that the column-averaged CO abundance decreases for higher disk masses and that the height below which CO freezes out increases with disk mass. This decreasing CO abundance means that the gas column density at \rgas\ needs to increase with disk mass in order to reach the same constant CO column density.

While this empirical correlation is evident in the models and can be understood qualitatively, it is difficult to reproduce it quantitatively. Appendix \ref{app: toy model} shows how this could be done using a toy model. It also shows that $R_{\tau_{\rm CO}=1}$, the radius where $^{12}$CO 2-1 becomes optically thin, is the more logical choice for such a model, rather than \rgas. However, while this toy model is able to show a correlation between $N_{\rm gas}(R_{\tau_{\rm CO}=1})$ and \mdisk, in practice the empirical relation between $N_{\rm gas}(\rgas)$ and \mdisk\ shown in Figure \ref{fig: Ngas-mdisk correlation} provides a much tighter correlation. In light of this we will use this empirical correlation throughout the rest of this work.

\subsection{Robustness of the correlation against varying disk parameters}
\label{sec: correlation -- dependencies}
\begin{figure*}
    \centering
    \includegraphics[width=\textwidth]{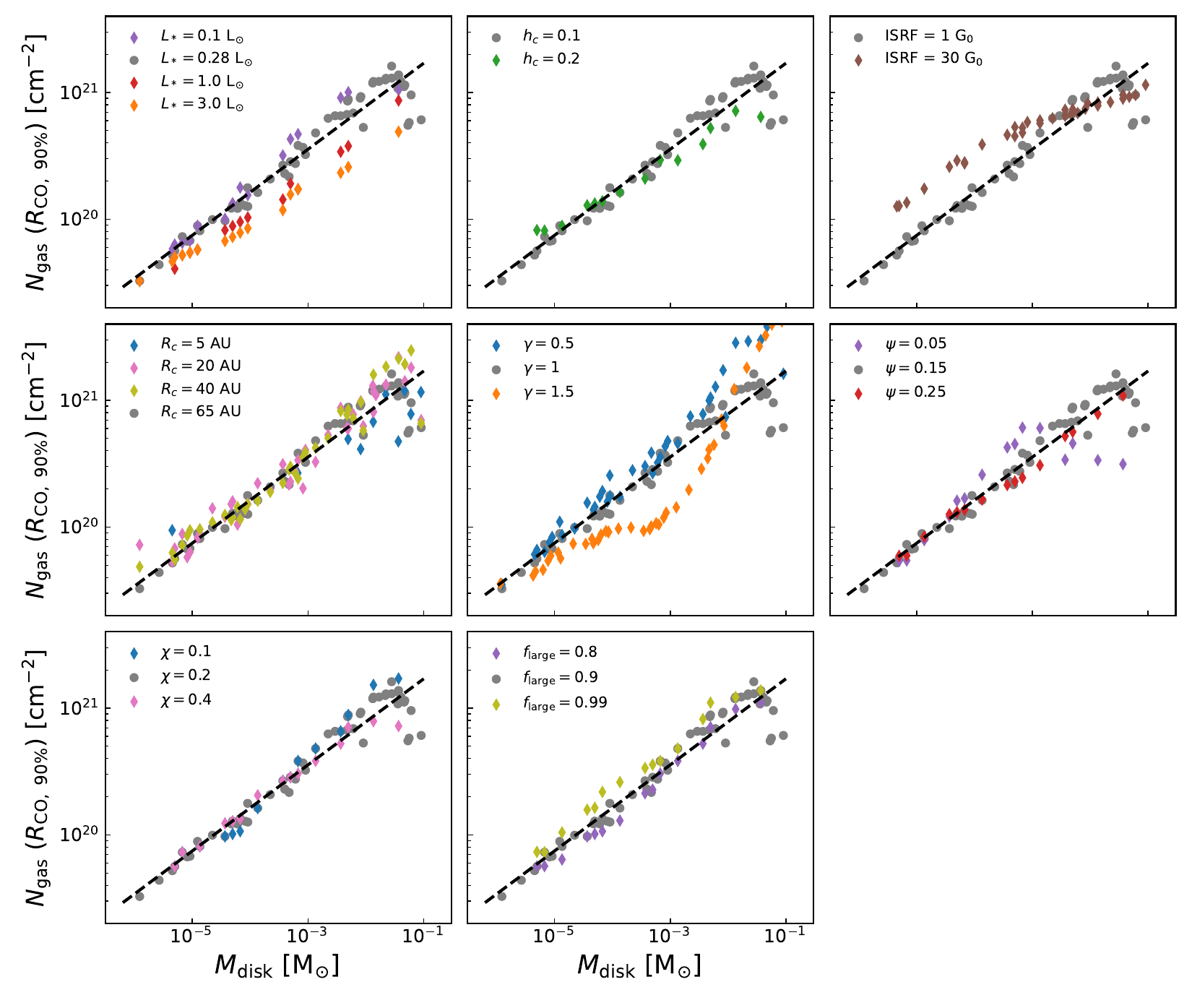}
    \caption{\label{fig: Ngas-mdisk correlation -- parameters} Correlation between $N_{\rm gas}(\rgas)$ and \mdisk\ examined for a wide range of disk and stellar parameters. From left to right, top to bottom the examined parameters are the stellar luminosity$(L_*)$, the scale height at \rc\ $(h_c)$, the external interstellar radiation field (ISRF), the characteristic size $(\rc)$, the slope of the surface density $(\gamma)$, the disk flaring angle $(\psi)$, the scale height reduction of the large grains $(\chi)$ and the fraction of large grains $(f_{\rm large})$. The gray points in each panel show the fiducial models shown in Figure \ref{fig: Ngas-mdisk correlation}. The black dashed line shows $N_{\rm gas}(\rgas)\propto\mdisk^{0.34}$. }
\end{figure*}

Figure \ref{fig: Ngas-mdisk correlation -- parameters} shows the correlation between $(N_{\rm CO}(\rgas))$ and \mdisk\ not only shows up for a single set of models but is unaffected by most disk parameters. The exceptions are the stellar luminosity, the strength of the external interstellar radiation field (ISRF) and the slope of the surface density profile. The stellar luminosity directly affects the temperature structure of disk. Increasing it moves the CO snow surface closer to the midplane. This increases the column averaged CO column at \rgas, which reduces the gas column needed to obtain the critical CO column density. 

Increasing the ISRF has two effects on the location of \rgas. Firstly, a larger CO column, and therefore also a larger gas column, is required to self-shield the CO against the stronger UV radiation field. Secondly, the external radiation will heat up the gas in the outer disk, which can thermally desorb CO ice back into the gas. This will increase the column averaged CO abundance, moving $N_{\rm gas}(\rgas)$ down again. The latter effect likely explains why for high disk mass both sets of models coincide again in Figure \ref{fig: Ngas-mdisk correlation -- parameters}.

Finally, models with a steeper surface density slope $(\gamma=1.5)$ have a much shallower exponential taper in the outer disk $(\Sigma_{\rm gas,outer}\propto\exp[-(R/\rc)^{2-\gamma}]$. Depending on the mass of the disk the CO emission in this taper can be partially optically thin. Inspection of the models shows that ones with $\gamma=0.5-1$ have $\tau\gtrsim1$ at \rgas, whereas models with $\gamma=1.5$ have $\tau\approx0.1$ at this radius. The presence of significant optically thin CO emission means \rgas\ no longer directly traces the radius where CO stops being able to self-shield. 
This is an important reason why $N_{\rm gas}(\rgas)$ scales with \mdisk\ (see Appendix \ref{app: toy model} for details). 

\subsection{Deriving an analytical expression for \rgas}
\label{sec: deriving analytical rgas}

If we fit the models presented in the previous section with a simple powerlaw between the gas column density at \rgas\ and the disk mass, we obtain
\begin{equation}
\label{eq: gas column disk mass relation}
N_{\rm gas}(\rgas)\equiv N_{\rm gas,crit} \approx 3.7\times10^{21} \left(\frac{M_{\rm gas}}{\mathrm{M}_{\odot}}\right)^{0.34}\ \mathrm{cm}^{-2}.   
\end{equation}
As discussed in the previous section, most disk parameters do not affect this powerlaw. Of the ones that do, only the stellar luminosity dependence can be readily included, as it only changes the slope and normalization of the powerlaw by a small factor. If we fit the stellar luminosity dependence of these two parts of our powerlaw, we obtain  
\begin{equation}
N_{\rm gas,crit} \approx 10^{21.27 - 0.53\log_{10} L_*} \left(\frac{M_{\rm gas}}{\mathrm{M}_{\odot}}\right)^{0.3 - 0.08\log_{10} L_*}\ \mathrm{cm}^{-2}.   
\end{equation}

As showed in the recent work by \cite{Toci2023} we can use this critical gas column density to obtain an analytical expression for \rgas. While \citet{Toci2023} left this critical value as a free parameter ($\Sigma_{\rm crit}$ in their notation), our models provide a quantitative estimate for this parameter.

Because the analytical solution contains a special function, Lambert-W function or product-log function, it is convenient to consider the case in which $\rgas \gg \rc$, i.e. that \rgas\ lies far into the exponential taper of the surface density profile. This case is more traceable and it is 
straightforward to show from Eqs. \eqref{eq: surface density} and \eqref{eq: gas column disk mass relation} that the observed outer radius scales with the logarithm of the disk mass 
\begin{align}
    \label{eq: exponent only}
    \frac{\mu m_H N_{\rm gas,crit}}{\Sigma_c} &\approx \exp\left[-\left(\frac{\rgas}{R_c}\right)^{2-\gamma}\right]\\
    \rgas &\overset{\gamma=1}{\approx} \rc\left[0.66\ln\mdisk - 2\ln\rc + const\right], \label{eq: approx scaling rgas}
\end{align}
where \mdisk\ and \rc\ are in units of \msun\ and au, respectively.

To first order the observed outer radius is thus expected to scale with the logarithm of the disk mass. Its dependence on \rc\ is more complex and will be explored in the next section.

If the surface density profile (Eq. \eqref{eq: surface density}) is inverted without any simplifying assumptions we obtain the following analytical prescription for \rgas\ as function of \mdisk, \rc\ (and $L_*$):\footnote{{Note that if the stellar luminosity dependence of $N_{\rm gas,crit}$ is included, the term in the square brackets of Equations \eqref{eq: full prescription}, \eqref{eq: simpler prescription - viscous} and \eqref{eq: simpler perscription - wind} becomes 
\begin{equation}
    [..] = 9.862\times10^{7 + 0.53\log_{10} L_*} \left(\frac{M_d}{\msun}\right)^{0.70 + 0.08 \log_{10} L_*} \left(\frac{\rc}{\rm au}\right)^{-2},
\end{equation}
where $L_*$ is in units of L$_{\odot}$.}}
\begin{equation}
    \label{eq: full prescription}
        R_{\rm CO,\ 90\%} = R_c \left(\frac{\gamma-\xi}{2-\gamma} W\left(\frac{2 -\gamma}{\gamma-\xi} \left[ 4.9\cdot10^7  \left(\frac{M_{\rm d}}{\mathrm{M}_{\odot}}\right)^{0.66} \left(\frac{\rm au}{R_c}\right)^{2} \right]
        ^{\frac{2-\gamma}{\gamma -\xi}}\right)  \right)^{\frac{1}{2-\gamma}}
\end{equation}
Here $W(z)$ is the Lambert-W function, or product-log function, specifically its principal solution $(k=0)$.

For common assumptions of a viscously evolving disk, i.e. $\gamma=1$ and $\xi=0$, the prescription reduces to
\begin{equation}
    \label{eq: simpler prescription - viscous}
        R_{\rm CO,\ 90\%} = R_c\times W\left( \left[ 4.9\times10^7  \left(\frac{M_{\rm disk}}{ \mathrm{M}_{\odot}}\right)^{0.66} \left(\frac{R_c}{\rm 1\,au}\right)^{-2} \right] \right)
\end{equation}

\noindent Similarly, for $\gamma=1$ and $\xi=0.25$ (the values of the fiducial models in this work)
\begin{equation} 
    \label{eq: simpler perscription - wind}
        R_{\rm CO,\ 90\%} = \frac{3 R_c}{4}\times W\left(\frac{4}{3} \left[ 4.9\times10^7  \left(\frac{M_{\rm disk}}{\mathrm{M}_{\odot}}\right)^{0.66} \left(\frac{R_c}{\rm 1\,au}\right)^{-2} \right]^{\frac{4}{3}}\right).
\end{equation}

\begin{figure}
    \centering
    \includegraphics[width=\columnwidth]{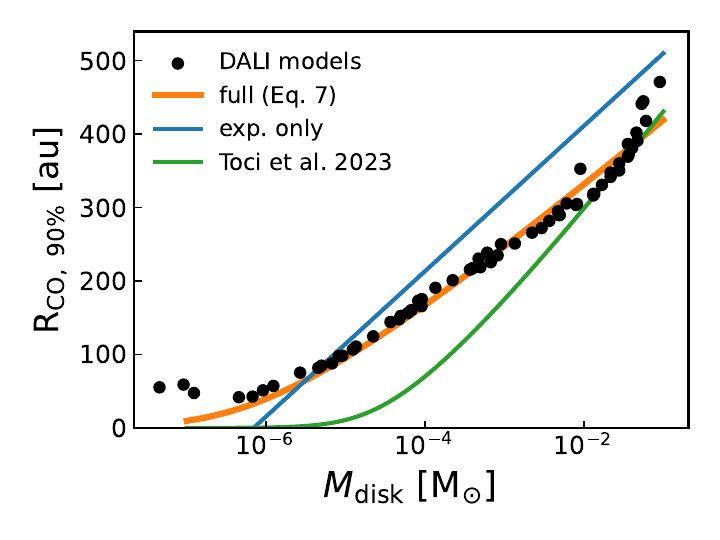}
    \caption{\label{fig: analytical Rco versus model} Comparison between the analytically calculated \rgas\ and the one obtained from the models using the fiducial disk parameters given in Table \ref{tab: model fixed parameters}. Black points show the model \rgas. The orange line shows \rgas\ calculated using Eq. \eqref{eq: full prescription}. The blue line shows \rgas\ calculated by approximating the surface density profile by its exponential taper (Eq. \eqref{eq: exponent only}). The green line shows the equivalent of expression for \rgas\ from \cite{Toci2023}, see Section \ref{sec: deriving analytical rgas} for details.  }
\end{figure}

Equation \eqref{eq: full prescription} allows us to analytically calculate \rgas\ from just \rc, \mdisk, $L_*$ and the slope of the surface density.  Before we use it, however, it is worthwhile to examine how well it reproduces the \rgas\ obtained from our disk models. Figure \ref{fig: analytical Rco versus model} shows this comparison for both the approximation that the dominant part of the surface density profile is its exponential taper (see Eq. \eqref{eq: exponent only}) and for the full derivation of an analytical \rgas\ (Eq. \eqref{eq: full prescription}).

The approximation of the surface density as just its exponential taper, as was proposed in e.g. \cite{Trapman2022}, captures the general trend of \rgas\ increasing with \mdisk, but does not match the exact shape of the mass dependence of \rgas. 
The \rgas\ calculated using Eq. \eqref{eq: full prescription} greatly improves the match, showing excellent agreement with the \rgas\ obtained from the disk models. Only for the very lowest and highest disk masses do we see a significant difference between the models and the analytical \rgas. Note that these are the same models where $N_{\rm gas}(\rgas)$ does not follow the powerlaw relation with \mdisk\ (see Figure \ref{fig: Ngas-mdisk correlation}). 

Figure \ref{fig: analytical Rco versus model} also shows the equivalent of the expression for \rgas\ presented by \cite{Toci2023}, who derive \rgas\ from where the surface density reaches a critical value  $\Sigma_{\rm crit} = \xi_{\rm CO}^{-1}\,2 \,m_H\,\hat{N}_{\rm CO}.$ The line shown here is for their adopted best values, $\xi_{\rm CO}=10^{-6}$ and $\hat{N}_{\rm CO} = 10^{16}\ \mathrm{cm}^{-2}$. Around a disk mass of $\mgas\approx 10^{-2} - 10^{-1}\ \msun$ agrees well with both the models and the analytical expression for \rgas\ from this work. In their work \cite{Toci2023} use a fiducial initial disk mass of 0.1\msun\ and evaluate the viscous evolution of \rgas\ between 0.1 and 3 Myr. Given the fact the mass of viscously evolving disks only decreases slowly over time ($\mdisk\propto^{-0.5}$ for $\gamma=1$) the disk masses covered in their work mostly lie in the $\mdisk\approx 10^{-2} - 10^{-1}\ \msun$ range where the models and the analytical expressions all agree.

\subsection{The link between \rgas\ and \rc}
\label{sec: rgas versus rc}

\begin{figure}
    \centering
    \includegraphics[width=\columnwidth]{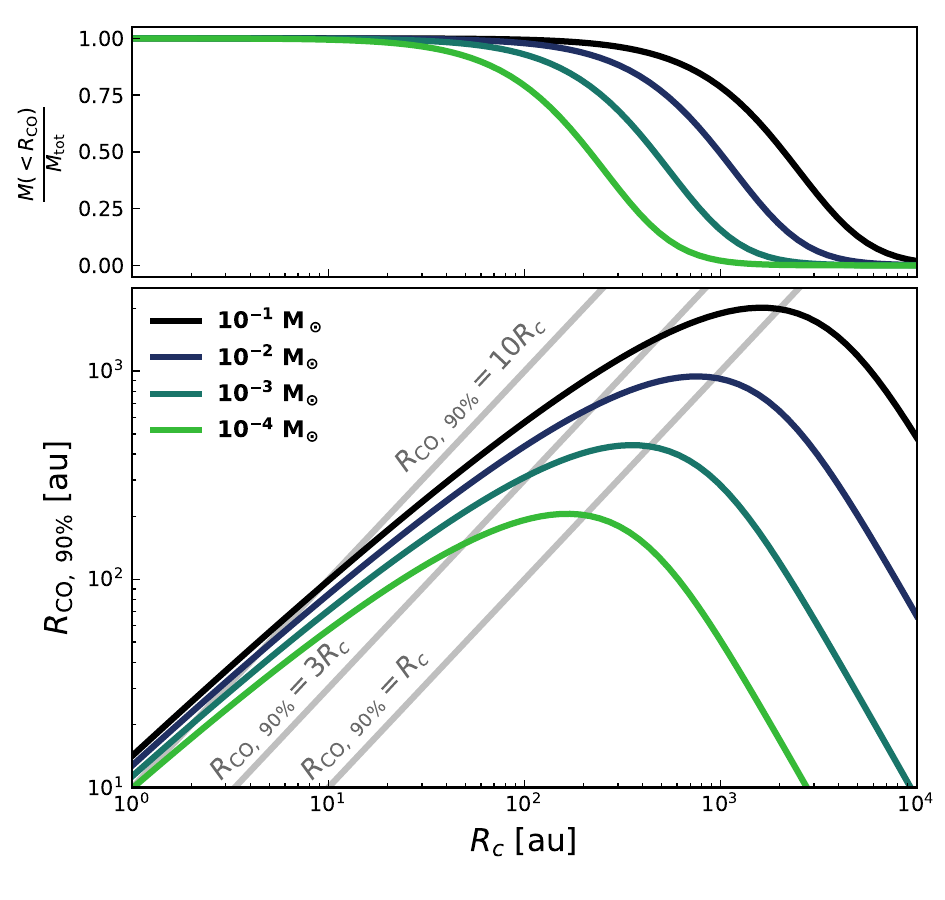}
    \caption{\label{fig: Rco vs Rc} \textbf{bottom panel:} Analytical \rgas\ calculated using Eq. \eqref{eq: full prescription} for a range of \rc. Colors show different disk masses. 
    \textbf{top panel:} Fraction of the total disk mass that is within \rgas, corresponding the solid line in the bottom panel.}
\end{figure}

\begin{figure}
    \centering
    \includegraphics[width=\columnwidth]{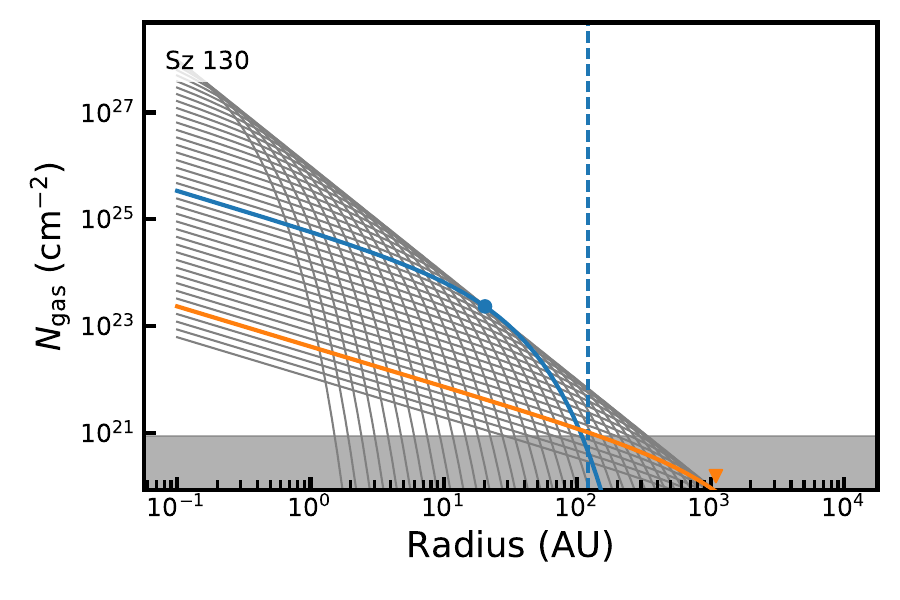}
    \caption{\label{fig: sigma profile fit} Gas surface density profiles calculated for different \rc\ but that all have the same total disk mass. The top of the gray shaded region shows the critical gas column density at \rgas\ for this disk. The vertical dashed line shows the observed \rgas. Two profiles with different \rc(marked by colored symbols) have this column density at \rgas.}
\end{figure}

Up to this point we have computed the observed radius \rgas\ for models where \rc\ was given. Observationally, however, we are interested in solving the opposite problem: for a given \rgas\ that was obtained from observations, what is the corresponding \rc? 
To that end, having vetted Equation \eqref{eq: full prescription} using our DALI models, we can now use it to study the relation between \rgas\ and \rc\ in disks. Figure \ref{fig: Rco vs Rc} shows \rgas\ as a function of \rc\ for four different disk masses using $\gamma=1$ and $\xi=0.25$. 
The shape of the curve shows that there are two values of \rc\ that can be inferred from a measurement of \rgas.
Figure \ref{fig: sigma profile fit} is a visualization of this, showing a set of example gas surface densities that all have the same total disk mass but a different \rc. Two profiles intersect with $N_{\rm gas}(\rgas)$ at \rgas: \rc\ = 20 au and \rc\ = 1000 au. The first \rc\ is much smaller than \rgas, meaning that \rgas\ lies in the exponential taper, while the other \rc\ that is larger than \rgas\ lies in the powerlaw part of the surface density. 
We should note however that while the "powerlaw"-\rc\ is a mathematical solution for \rgas\ it is also an extrapolation for Eq. \ref{eq: full prescription} beyond the domain where it was tested. None of the DALI models examined in this work have $\rc > \rgas$ and it is entirely possible that disks with such a disk structure, likely those with a very low disk mass, do not follow the $N_{\rm gas}(\rgas)-\mdisk$ correlation on which Eq \eqref{eq: full prescription} is build. 

Interestingly, the curves in Figure \ref{fig: Rco vs Rc} also imply that for a given disk mass there is a maximum observed disk size, where \rgas\ is equal to \rc. Increasing \rc\ beyond this point decreases \rgas\ as a large fraction of the disk mass ($\gtrsim 50 \%$, see the top panel of Figure \ref{fig: Rco vs Rc}) now exists as low surface density material below the CO photodissociation threshold. A demonstration of this effect can be seen in the evolution of \rgas\ for a low mass viscously evolving disk. As can be seen in \cite{Trapman2020} (e.g. their Figure 3), the \rgas\ of a low mass, high viscosity disk first increases with time until the rapid viscous expansion lowers the surface density to the point where the CO photo-dissociation front starts moving inward, resulting in \rgas\ now decreasing with time.

The existence of a maximum \rgas\ for each disk mass also suggests that \rgas\ places a lower limit on the disk mass. By taking the derivative of \rgas\ to \rc\ and setting it to zero, this minimum disk mass can be written as (for the derivation, see Appendix \ref{app: deriving minimum mass}) 
\begin{equation}
\label{eq: mass lower limit}
\mdisk \gtrsim 1.3\times10^{-5} \left( \frac{\rgas}{\rm 100\ au} \right)^{3}\ \mathrm{M_{\odot}}
\end{equation}

It should be kept in mind however that this disk mass has been derived by assuming a surface density profile and fitting it through a single point ($N_{\rm gas}$ at \rgas). Its accuracy therefore depends on how well this surface density profile matches the actual surface density of protoplanetary disks.

\section{Discussion}
\label{sec: discussion}

\subsection{Extracting an estimate of \rc\ from observed \rgas.}
\label{sec: extracting R_c}

\begin{figure}[htb]
    \centering
    \includegraphics[width=\columnwidth]{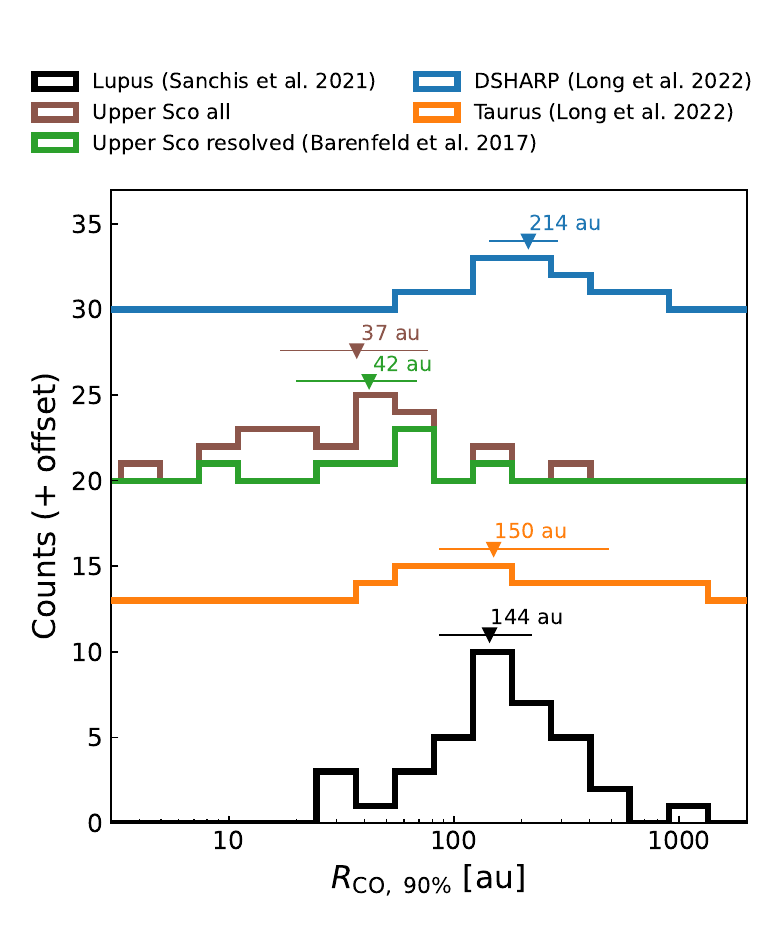}
    \caption{\label{fig: observed disk size distribution} Distribution of observed disk sizes \rgas\ for Lupus (black), Upper Sco (brown/green), Taurus (blue) and DSHARP (orange). The resolved Upper Sco sample shown here includes only the seven well resolved sources, while the full sample includes all sources where $^{12}$CO was detected (for details, see \citealt{barenfeld2017}). The triangles denote the median \rgas\ and the horizontal line shows the 25$^{\rm th}$ and 75$^{\rm th}$ quantile of the \rgas\ distribution of each region. }
\end{figure}

In the previous Section we showed that Equation \eqref{eq: full prescription} provides the link between \rgas\ and \rc\ based on \mdisk. Leveraging this equation we can derive \rc\ from the observed disk sizes that have now been measured from $^{12}$CO emission for a large number of disks distributed over several star forming regions\footnote{Note that for the DSHARP sample we limit ourselves to the sources without severe clouds contamination, see \cite{Long2022} for more details.} (e.g. \citealt{barenfeld2017,ansdell2018,Sanchis2021,Long2022}, see Table \ref{tab: observed sample} and Figure \ref{fig: observed disk size distribution}). Before we continue there are two things that should be kept in mind.
The observations from which these sizes are measured are shallow, which means that the uncertainties on most \rgas\ are large, up to 30 \% (see \citealt{Sanchis2021}). Another good example of this are the observations of disks in Upper Sco, where \cite{barenfeld2017} detected $^{12}$CO 3-2 in 23 of the 51 continuum detected sources, but from fitting the CO visibilities was only able to provide well constrained gas disk sizes (i.e. statistically inconsistent with 0) for 7 disks in the sample. So when deriving \rc\ we have to take the uncertainties on \rgas\ into account.

Inverting Eq. \eqref{eq: full prescription} also requires the disk gas mass, which is a difficult quantity to measure. Gas masses derived from CO isotopologue emission are found to be low ($\lesssim 1 M_{\rm jup}$, see e.g. \citealt{ansdell2016,miotello2017,Long2017}). However, there are large uncertainties on the CO abundance in disks (e.g. \citealt{Favre2013,Schwarz2016,Zhang2019,Zhang2020b,Trapman2022b}). We will therefore make the assumption that all disks have a gas-to-dust mass ratio of 100. For the disks where the gas mass is measured using HD the gas-to-dust mass ratio seems approximately 100, although this is only for a few disks in a very biased sample. 
New observations from the ALMA survey of Gas Evolution in Protoplanetary disks (AGE-PRO) will allow us to overcome this hurdle by measuring accurate gas masses for 20 disks in Lupus and Upper Sco, using \nhp\ observationally constrain their CO abundance (see \citealt{Trapman2022b} for details).
We will discuss the assumption of a single gas-to-dust mass ratio later in this section.

\newtext{The details for our approach of obtaining \rc\ from \rgas\ can be found in Appendix \ref{app: derived rc}}. Before continuing to the \rc-distributions of our various samples, let us first examine the computed \rc\ for five well known disks that have been previously studied in detail using thermochemical models that reproduce, among a number of other observables, the observed extent of CO and its isotopologues: TW Hya, DM Tau, IM Lup, AS 209 and GM Aur \citep{Kama2016,Zhang2019,Zhang2021MAPS,Schwarz2021MAPS}. For three of the five disks, DM Tau, IM Lup and TW Hya the simple estimate in Table \ref{tab: observed sample} roughly agrees with the \rc\ in the more detailed studies. Not so for GM Aur and AS 209 however, which have estimated \rc\ that are much smaller than the literature values. For AS 209, the difference in \rc\ can be traced back to the fact that the disk mass used here (i.e. $100\times\mdust$) is $\sim10\times$ larger than the one derived by \cite{Zhang2021MAPS}. For GM Aur it is harder to identify a similar cause. 
It should be noted though that fitting \rc\ was not the primary goal of the previous studies discussed here. These detailed models reproduce the observations for the given \rc, but due to the complexity of the fitting it is hard to determine how unique these values of \rc\ are.

\begin{figure}
    \centering
    \includegraphics[width=\columnwidth]{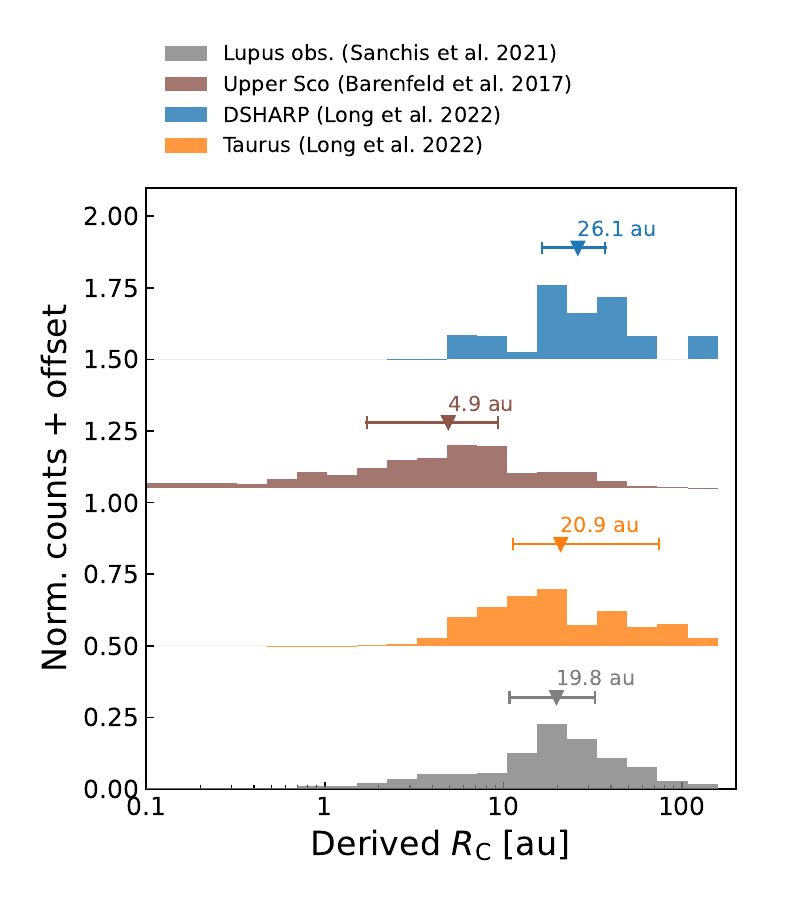}
    \caption{\label{fig: distribution of Rc} Derived distributions of \rc\ for four disk samples: Lupus (gray; \citealt{ansdell2018,Sanchis2021}), Upper Sco (brown; \citealt{barenfeld2017}), Taurus (orange; \citealt{Long2022}) and DSHARP (blue; \citealt{Andrews2018,Long2022}). The triangles denote the median \rc\ and the horizontal line shows the 25$^{\rm th}$ and 75$^{\rm th}$ quantile of the \rc\ distribution of each region. See Table \ref{tab: observed sample} for the \rc\ of individual disks. }
\end{figure}

To examine and compare the distributions of \rc\ in different star-forming regions, we sum up the distributions of \rc\ for individual sources in each region and normalize the resulting distribution. Figure \ref{fig: distribution of Rc} shows the normalized distribution of \rc\ of Lupus, Upper Sco, Taurus and the DSHARP sample. Lupus and Taurus have similar median \rc, 19.8$^{+12.8}_{-9}$ au and 20.9$^{+54.4}_{-9.6}$ au for the two regions respectively, while the DSHARP sample has a slightly larger median $\rc = 26.1^{+12.1}_{-9.7}$ au. Here the uncertainties denote the 25\% and 75\% quantile of the distribution. The clear outlier is Upper Sco with a median $\rc = 4.9^{+4.4}_{-3.2}$. This is a surprising find given the age difference between Lupus/Taurus ($\sim1-3$ Myr, e.g. \citealt{comeron2008}) and Upper Sco ($\sim5-11$ Myr, e.g. \citealt{Preibisch2002,Pecaut2012}). Figure \ref{fig: distribution of Rc} thus shows a decrease in \rc\ with time, which does not match with predictions from either of the predominant theories of disk evolution. Viscously evolving disks are expected to grow over time, with \rc\ increasing with age. Conversely, disks evolving under the effect of magneto-hydrodynamical disks winds are expected to have an \rc\ that is constant with time. \newtext{Even a combination of viscous and MHD wind-driven evolution would be hard pressed to explain the decrease of \rc, given the inability of both components to explain the observed decrease in \rc.} A potential cause for the systematically smaller \rc\ in Upper Sco is the environment in which these disks find themselves, specifically their proximity to the nearby Sco-Cen OB association. Ultraviolet radiation from these O- and B-stars could have truncated the disks (e.g. \citealt{Facchini2016,Haworth2017,Haworth2018,Winter2018}), resulting in a different evolutionary path compared to the disks in the more quiescent Lupus and Taurus star-forming regions. Note that in the case of truncated disks the \rc\ values derived here for Upper Sco should be viewed with caution, as they are derived under the assumption of a tapered powerlaw surface density profile, an assumption which is no longer valid in this case. \newtext{We reserve a more comprehensive analysis of the effect of external photo-evaporation on \rgas\ in Upper Sco for a future work.}

\subsection{Caveats and limitations}
\label{sec: caveats}

When comparing median \rc\ of different regions in Section \ref{sec: extracting R_c} there are several factors that we should keep in mind.
The first is that none of these samples are complete. Due to limited sensitivity of the observations the faintest and most compact sources are likely not detected and thus not included in the sample. The inclusion of these sources would decrease the median \rc\ if they are compact, but without deep observations we cannot rule out the existence of large, low surface brightness disks that would increase the median \rc. 

\newtext{Similarly, the binarity of the samples should be considered. Binaries can truncate the disk and, more generally, disks in multiple systems evolve differently than those around single stars (see, e.g. \citealt{Kraus2012,RosottiClarke2018,Zagaria2021,Zagaria2022}). Indeed, there is some suggestion that Upper Sco has a higher binary fraction than Lupus (e.g. \citealt{Barenfeld2019,Zurlo2021,Zagaria2022}). However, this should not be taken at face value, as our samples are not complete and the binarity surveys are not homogeneous (see appendix A of \citealt{Zagaria2021} for an extensive discussion on this). A homogenous study of disk multiplicity is needed to conclusively show its effect on disk sizes.}

There is also a difference in methodology that needs to be considered. The \rgas\ in Lupus, Taurus and DSHARP were all measured from the integrated intensity map of the CO emission. As mentioned above, the \rgas\ of Upper Sco are measured from a CO intensity profile that was fitted to the visibilities (see \citealt{barenfeld2017}). It is possible that this introduces some systematic effect that results in lower values for \rc\ in Upper Sco. \newtext{These observed \rgas\ are then compared to the noiseless, high resolution synthetic CO observations of our models, which are most akin to the DSHARP observations (see, e.g., Section 3.1.2 in  \citealt{Sanchis2021} for a detailed discussion how higher resolution and/or sensitivity affects the measurement of \rgas).  }
It is also worth pointing out that the Upper Sco gas disk sizes are measured from the \co\ $J=3-2$ line rather than the $J=2-1$ line used for the other regions, but models show that has only a small $(\lesssim 10\%
)$ effect on \rgas\ (e.g. \citealt{Trapman2019}).
The forthcoming AGE-PRO observations will test this possibility by consistently measuring gas disk sizes for a carefully selected sample of disks in Lupus and Upper Sco.

The assumption of a single gas-to-dust mass ratio for all sources irrespective of their age is also likely to be incorrect. Dust evolution models show that the gas-to-dust ratio increases with age as more of the dust mass is converted into larger bodies that either drift inward and are accreted onto the star (e.g. \citealt{Birnstiel2012}) or form planetesimals that do not emit at millimeter wavelengths and are thus unaccounted for in our dust masses (e.g. \citealt{Pinilla2020}). This would however only increase the difference between Upper Sco and the younger regions, as to explain the same \rgas\ with a higher mass disks requires a smaller \rc. Using a lower gas-to-dust mass ratio for Upper Sco would move the median \rc\ closer to the values of Lupus and Taurus. However, Figure \ref{fig: Rco vs Rc} shows that the effect of changing the disk mass is small. To produce an \rgas\ of, for example, 60 au requires a \rc\ of $\approx5$ au if the disk has a mass of $\mdisk=0.1\ \msun$, which increases to $\rc\approx10$ au for a disk that is three order of magnitude less massive ($\mdisk=10^{-4}\ \msun$).  

Another source of uncertainty is the global CO abundance in the disk. The processes that have been proposed for removing CO from the gas in disks (beyond CO freeze-out and photo-dissociation) are expected to operate on Myr timescales (e.g. \citealt{Krijt2018,Krijt2020,Yu2017,Bosman2018b}), which is corroborated by observations (e.g. \citealt{Zhang2020b}). In addition to differences between individual sources we can thus expect a trend of lower CO abundances with age. Observations of N$_2$H$^+$ of two disks in Upper Sco suggest that this is indeed the case (see \citealt{Anderson2019}). If the overall CO abundance in the disk is lower the gas column at \rgas\ needs to be larger to build up a CO column capable of self-shielding against photo-dissociation. Given that the total disk mass is fixed the derived \rc\ will have to increase to explain the same \rgas\ with a lower CO abundance. 

Quantifying the effect on \rgas\ depends on the exact physical and/or chemical processes responsible for removing the CO from the gas, but also, maybe even more importantly, on how well-mixed the disk is vertically. If vertical mixing is inefficient CO could be removed from the midplane, as traced by \xco\ and \cyo, while the upper layers of the disk from which \co\ emits remain unaffected. In this case, \rgas\ would not be significantly affected by a decrease in CO abundance (see \citealt{Trapman2020}).

Conversely, if the disk is well-mixed vertically the CO abundance in the \co\ emitting layer will also lower than currently assumed. \cite{Trapman2022} showed that this, coupled with the relatively poor brightness sensitivity of the shallow ALMA disk surveys, can significantly reduce the observed value of \rgas. Accounting for this fact would bring the characteristic radii of Upper Sco closer to those of disks in Lupus and Upper Sco. Recent work by \cite{Zagaria2023} arrived at a similar conclusion.
We should also note that CO depletion factor is seen to vary with radius \citep{Zhang2019,Zhang2021MAPS}, which complicates extrapolating the CO abundance of the bulk of the gas to the region in the outer disk that is most relevant for setting \rgas. 

\emph{The shape of the surface density in the outer disk} is an important part in the analytical relation between \rgas\ and \rc\ presented in this work (see also Appendix \ref{app: toy model}). Most notably, our models assume that the surface density follows an exponential taper. While this assumption is well grounded in theory, observational constraints on the surface density in the outer part of disks are sparse (e.g. \citealt{Dullemond2020}). Figure \ref{fig: Ngas-mdisk correlation -- parameters} gives us some idea about in what way the surface density must be different to nullify the $N_{\rm gas}(\rgas)-\mdisk$ relation. If $\gamma$ is decreased, in which case the exponential taper becomes steeper and the surface density starts to approach a truncated powerlaw, the $N_{\rm gas}(\rgas)-\mdisk$ relation is retained. This suggests that the relation should be there for disks where the surface density drops of steeply, whereas for disks with a shallow surface density profile in the outer disk the relation will no longer hold and the analytical expression for \rgas\ should not be used. However, we should remain cautious when extrapolating from our ``$\gamma$-models''. By construction, a steeper exponential taper (i.e. small $\gamma$) corresponds to a flatter powerlaw at small radii and vice versa for large $\gamma$. In the end it is always prudent to use tailored models for disks with noticeably, or expected, different surface density profiles rather than use a generalized model.
\newtext{In a similar vein, substructures in the gas and radial variations in the gas-to-dust mass ratio could affect \rgas. However, as \rgas\ is measured from the optically thick $^{12}$CO emission these structures would need to meaningfully change the temperature structure in the $^{12}$CO emitting layer to affect the $^{12}$CO emission profile and therefore \rgas. \cite{Law2021aMAPS} showed that the high resolution $^{12}$CO observations show comparatively little substructure in contrast to more optically thin CO isotopologues and the dust. However, if a substructure near \rgas\ were to locally change the temperature structure and thereby change the location of the CO snow surface it would likely break the $N_{\rm gas}(\rgas)$-\mdisk\ correlation on which the analytical equation of \rgas\ is build. That being said, most substructures are found much closer to the star, far away from \rgas, meaning their effect on \rgas\ is likely minimal.}

Similarly, the \emph{temperature structure of the disk and its vertical structure} or more precisely, how much of the CO column is frozen out is a key link in the correlation of N$_{\rm gas}(\rgas)$ and \mgas, as demonstrated by the tight correlation between $z_{\rm freeze} (\rgas)$ and \mgas. We have explored the parameters that predominantly affect the temperature structure in our models. From the observational side, several recent studies have used high resolution ALMA observations of CO to map the radial and vertical temperature structures of disks (e.g. \citealt{Pinte2018,Law2021bMAPS,Law2022,Paneque-Carreno2023}). Temperature structures computed with models similar to the ones in this work have been found to match these observationally constraints (e.g. \citealt{Zhang2021MAPS}). However, the number of disks with good observational constraints on their 2D temperature structures is still limited and, due to requirement of deep, high resolution observations, biased to large disks. There is therefore still the possibility that our models do not accurately describe the temperature structure of all disks, in which case it is very likely that the analytical expression for \rgas\ presented in this work will no longer hold. 

\section{Conclusions}
\label{sec: conclusions}

In this work we have presented an empirical relation between the gas column density measured at the observed gas outer radius $(N_{\rm gas}(\rgas))$ and the mass of the disk \mdisk. Using this relation we provided simple prescriptions for conversions of \rc\ to \rgas\ and from \rgas\ to \rc (Eq. \ref{eq: full prescription}).
Our main take-away points are:
\begin{itemize}
    \item Using thermochemical models, we found an empirical correlation between the gas column density at the observed gas disk size \rgas\ and the mass of the disk: $N_{\rm gas}(\rgas) \approx 3.7\times10^{21}(\mdisk/\msun)^{0.34}\ \mathrm{cm}^{-2}$. Importantly, this correlation does not significantly depend on other disk parameters.
    \item Following \cite{Toci2023} we used this empirical correlation to provide an analytical prescription of \rgas\ that only depends on \rc\ and \mdisk. This analytical prescription is able to reproduce \rgas\ from thermochemical models for a large range of \mdisk\ and \rc. 
    \item Exploring the analytical prescription of \rgas\ reveals a maximum \rgas\ for a given \mdisk\ that is independent of \rc (Eq. \ref{eq: mass lower limit}). It also shows that for a given \mdisk\ any \rgas\ can be obtained with two different values of \rc\ ($\rc \ll \rgas$ or $\rc \gg \rgas$).  
    \item Using the observed \rgas\ and $\mgas=100\times\mdust$ we derived \rc\ for four samples of disks in Lupus, Upper Sco, Taurus and DSHARP. We find that Lupus and Taurus have similar median \rc, 19.8 and 20.9 au respectively, and the DSHARP disks are slightly larger ($\rc = 26.1$). Surprisingly, the disks in Upper Sco are significantly smaller, with a median $\rc=4.9$ au. This decrease in $\rc$ for the older Upper Sco region goes against predictions of both viscous and wind-driven evolution.
\end{itemize}

\begin{acknowledgements}
\newtext{We thank the referee for their valuable feedback, which helped to improve the quality of this manuscript}
L.T. and K. Z. acknowledge the support of the NSF AAG grant \#2205617.
B.T. acknowledges the support by the Programme National “Physique et Chimie du Milieu Interstellaire” (PCMI) of CNRS/INSU with INC/INP and co-funded by CNES.
GR acknowledges support from the Netherlands Organisation for Scientific Research (NWO, program number 016.Veni.192.233), from an STFC Ernest Rutherford Fellowship (grant number ST/T003855/1) and is funded by the European Union (ERC DiscEvol, project number 101039651). Views and opinions expressed are however those of the author(s) only and do not necessarily reflect those of the European Union or the European Research Council Executive Agency. Neither the European Union nor the granting authority can be held responsible for them.
All figures were generated with the \texttt{PYTHON}-based package \texttt{MATPLOTLIB} \citep{Hunter2007}. This research made use of Astropy,\footnote{http://www.astropy.org} a community-developed core Python package for Astronomy \citep{astropy:2013, astropy:2018} and SCIPY \citep{scipy,corless1996lambertw}.
\end{acknowledgements}

\bibliographystyle{aasjournal}
\bibliography{references}

\begin{appendix}

\section{A toy model for analytically deriving the observed CO outer radius}
\label{app: toy model}

Section \ref{sec: correlation -- first look} showed a clear correlation between the gas column density measured at \rgas, the radius that enclosed 90\% of the $^{12}$CO $J=2-1$ emission, and the total mass of the disk \mdisk. It also showed a similarly tight correlation between the height of the CO snow surface as measured at \rgas\ and \mdisk, giving a hint as to the origin of the first correlation. Here we will set up a simple toy model of the CO abundance in protoplanetary disks, link it to the resulting CO emission, and show how it can produce a correlation between the column density at the outer radius and the disk mass. 

\subsection{Concept and assumptions}
\label{sec: concept and assumptions}

Starting from the observations, it is common to use $^{12}$CO rotational emission to measure the size of protoplanetary disks. Low $J$ lines of CO become optically thick already at small column densities, making CO emission bright and easy to detect out to disk large radii. The transition from optically thick to optically thin CO emission thus occurs in the outer part of the disk, where the surface density likely declines steeply with radius. This is indeed the case if the surface density follows an exponential taper, but one should keep in mind that observational constraints of the shape of the surface density in the outer disk are very limited (see, e.g. \citealt{Cleeves2016,Dullemond2020}). Given that the density is low here, we can expect only a small contribution of the optically thin CO emission to the total CO flux. In other words, we expect that most, if not all, of the CO emission is optically thick. 

At the same time, we know that CO will become photo-dissociated in the outer disk. The exact CO column density required to self-shield against photo-dissociation depends somewhat on the molecular hydrogen column density and temperature, but in general the threshold is taken to be a CO column density of a few times $10^{15}\ \mathrm{cm}^{-2}$ (see, e.g.  \citealt{vanDishoeckBlack1988, Visser2009}). 
It is common to assume that the radius at which the CO line emission becomes optically thin coincides with the radius at which CO stops being able to self-shield, i.e., that the CO emission disappears beyond this point (e.g. \citealt{Trapman2019,Toci2023}). In this case we can give a simple description of the CO radial emission profile:

\begin{equation} 
\label{eq: approximate intensity}
  I_{\rm CO}(R) = 
    \begin{cases}
  T_0 \left(\frac{R}{R_0}\right)^{-\beta} & \text{if $N_{\rm CO}(R)\geq a\times10^{15}\ \mathrm{cm}^{-2}$}  \\
  0 & \text{otherwise}
    \end{cases}
\end{equation}
where $T_0 (R/R_0)^{-\beta}$ describes the temperature profile of the CO emitting layer as a simple powerlaw and $a$ is a constant of order unity.

Under the these simplifying assumptions, the radius that encloses 100 \% of the CO flux would be the radius where we reach $N_{\rm CO}(R)\approx a\times10^{15}\ \mathrm{cm}^{2}$. Note that definition commonly used in observations to measure gas disk sizes, i.e. \rgas, the radius that encloses 90\% of the flux, is very closely related to the 100\% radius (see, e.g., appendix F in \citealt{Trapman2019}): 
\begin{align}
    R_{\rm CO, 100\%} &= 0.9^{\frac{1}{\left(2-\beta\right)}}\rgas\\
                      &\approx 0.93\times\rgas\ \text{for}\ \beta=0.5
\end{align}

However, as we will discuss further on in this section, this small difference has a meaningful impact on the $N_{\rm gas}(\rgas)-\mdisk$ relation discussed in the main body of this work. For the rest of the derivation we will therefore use $R_{\rm CO, 100\%} = R_{\tau_{\rm CO} =1}\equiv R_{\tau}$ rather than \rgas.

The relation between the CO column density and the H$_2$ column density depends on the column averaged CO abundance. The zeroth order assumption would be that the CO abundance is a constant $10^{-4}$, where all of the available carbon is locked up in the gas. However, this ignores the fact that the disk becomes colder towards the midplane, causing the CO to freeze out and thus lowering the local CO abundance. Similarly, photo-dissociation will decrease the CO abundance in the uppermost layer of the disk. These two processes confine CO to a so-called warm molecular layer, first introduced as a concept by \citealt{Aikawa2002}. As a result, the column averaged CO abundance will be lower than $10^{-4}$. 

Given that most of mass in the column is concentrated towards the midplane we can, to first order, ignore the decrease in CO abundance due to photo-dissociation and write the vertical CO abundance profile as a simple step function

\begin{equation} 
\label{eq: parameteric CO}
  x_{\rm CO}(R,z) = 
    \begin{cases}
  0 & \text{if $z\leq z_{\rm freeze}(R)$}  \\
  x_{\rm CO,\ peak} & \text{if $z> z_{\rm freeze}(R)$},
    \end{cases}
\end{equation}
where $z_{\rm freeze}(R)$ describes the height of the CO ice-surface, which is approximately equivalent to $T_{\rm gas}(R,z_{\rm freeze}) = 20\ \mathrm{K}$ and we assume that $x_{\rm CO,\ peak} = 10^{-4}$.

In principle obtaining $z_{\rm freeze}(R)$ requires computing the 2D temperature structure of the disk. This can be done by assuming that $T_{\rm gas} \approx T_{\rm dust}$, a reasonable assumption for the area of interest here, and computing $T_{\rm dust}(r,z)$ by solving the radiative transfer equation (e.g.,
\citealt{RADTRAN,radmc3d,BrinchHogerheijde2010}). Alternatively, the temperature structure can be measured from optically thick emission lines (e.g., \citealt{Dartois2003, Dullemond2020,Law2021bMAPS,Law2022}). Here we will keep using $z_{\rm freeze}(R)$ until later in the derivation.

The vertical density distribution resulting from isothermal hydrostatic equilibrium is given by a Gaussian (e.g. \citealt{ChiangGoldreich1997})
\begin{equation}
\label{eq: vertical gaussian}
    \rho_{\rm gas} = \frac{\Sigma(r)}{\sqrt{2\pi}H(r)} \exp\left[-\frac{1}{2}\frac{z^2}{H(r)^2}\right],
\end{equation}
where $H(r)$ is the height of the disk.

To obtain the CO column density of our simple two-part CO abundance model (Eq. \eqref{eq: parameteric CO}) we need to find the column density above $z_{\rm freeze}$
\begin{align}
\label{eq: upper column}
N_{\rm gas}(r) &= \frac{\Sigma_{\rm gas}(r)}{\mu m_H}\\
N_{\rm > z_fr}(r) &= \int_{z_{\rm freeze}}^{\infty} \frac{\Sigma_{\rm gas}(r)}{\mu m_H \sqrt{2\pi} H(r)}\exp\left[-\frac{1}{2}\frac{z^2}{H(r)^2}\right] \mathrm{d}z\\
                  &= N_{\rm gas}(r) \frac{1}{\sqrt{2\pi} H(r)} \int_{z_{\rm freeze}}^{\infty} \exp\left[-\frac{1}{2}\frac{z^2}{H(r)^2}\right]\mathrm{d}z\\
                  &\stackrel{t=z/\sqrt{2}H}{=} \frac{N_{\rm gas}(r)}{\sqrt{\pi}} \int_{z_{\rm freeze}/\sqrt{2}H}^{\infty} \exp\left[-t^2\right]\mathrm{d}t\\
                  &= \frac{N_{\rm gas}(r)}{2} \left[1 - \mathrm{erf}\left(\frac{z_{\rm freeze}(r)}{\sqrt{2}H(r)} \right)\right].
\end{align}
Here $N_{\rm gas}$ is the gas column density, $\mu$ is the mean molecular weight, $m_H$ is the hydrogen atomic mass and $\mathrm{erf}$ is the error function.
This allows us to write out the CO column density above $z_{\rm freeze}$ (see eq. \eqref{eq: parameteric CO})

\begin{align}
\label{eq: CO column}
N_{\rm CO} &= x_{\rm CO} N_{\rm z > z_{\rm freeze}}\\
           &= \frac{x_{\rm CO,peak}}{2} N_{\rm gas}(r) \left[1 - \mathrm{erf}\left(\frac{z_{\rm freeze}(r)}{\sqrt{2}H(r)} \right)\right],
\end{align}.

Using the gas surface density instead of the gas column density $N_{\rm gas}(r)$, Equation \eqref{eq: CO column} becomes
\begin{equation}
\frac{2 N_{\rm CO}}{x_{\rm CO,peak}} = \frac{\Sigma_{\rm gas}(r)}{\mu m_H} \left[1 - \mathrm{erf}\left(\frac{z_{\rm freeze}(r)}{\sqrt{2}H(r)} \right)\right]
\end{equation}

We recall that in our toy model $R_{\tau}$ coincides with the radius where the CO column density is the critical CO column density needed for CO self-shielding $(N_{\rm CO}(R_{\tau} = N_{\rm CO,\ crit})$. We can then derive an expression for $R_{\tau}$ from the previous equations as:

\begin{align}
\label{eq: numerical equation}
\frac{2\mu m_H N_{\rm CO}}{x_{\rm CO,peak}} &\equiv2\Sigma_{\rm CO,crit} = \Sigma_c \left[1 - \mathrm{erf}\left(\frac{z_{\rm freeze}(R_{\tau})}{\sqrt{2}H(R_{\tau})} \right)\right]\\   
        &\times\left(\frac{R_{\tau}}{\rc}\right)^{-\gamma+\xi}\exp\left[-\left(\frac{R_{\tau}}{\rc}\right)^{2-\gamma}\right]\\
\frac{2 \Sigma_{\rm CO,crit}}{\Sigma_c} &\equiv \hat{\Phi} = \left[1 - \mathrm{erf}\left(\frac{z_{\rm freeze}(R_{\tau})}{\sqrt{2}H(R_{\tau})} \right)\right]\\   
        &\times\left(\frac{R_{\tau}}{\rc}\right)^{-\gamma+\xi}\exp\left[-\left(\frac{R_{\tau}}{\rc}\right)^{2-\gamma}\right]. \label{eq: numerical eq1}
\end{align}

A solution for a similar equation without the CO freeze-out term in the square brackets was recently presented by \cite{Toci2023}. 
Here we follow their work by introducing the shorthands $\Sigma_{\rm CO,crit}$ and $\hat{\Phi}$\footnote{For direct comparison with \cite{Toci2023}: $\Sigma_{\rm crit,toci+2022} = \Sigma_{\rm CO,crit}/[..]$ and $\Phi_{\rm toci+2022} = 0.5\hat{\Phi}/[..]$, where [..] is the term in square brackets in Equation \eqref{eq: numerical equation}.}.

With the introduction of the CO freeze-out term Equation \ref{eq: numerical eq1} can no longer be solved analytically. However if the vertical density and temperature structure are known and prescriptions for $H(R_{\tau})$ and $z_{\rm freeze}(R_{\tau})$, or more accurately $z_{\rm freeze}(R_{\tau})/H(R_{\tau})$, can be provided the equation can be solved numerically. 

\begin{figure*}
    \centering
    \begin{minipage}{0.49\textwidth}
        \includegraphics[width=\textwidth]{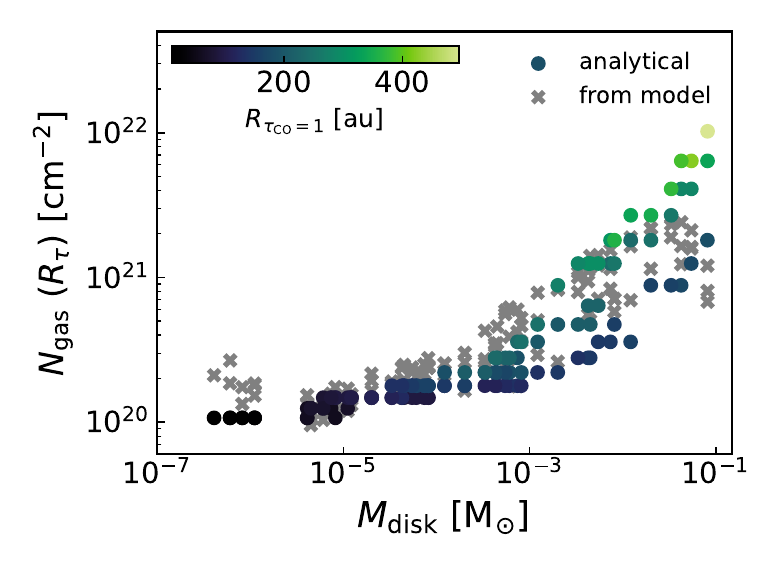}
    \end{minipage}
    \begin{minipage}{0.49\textwidth}
        \includegraphics[width=\textwidth]{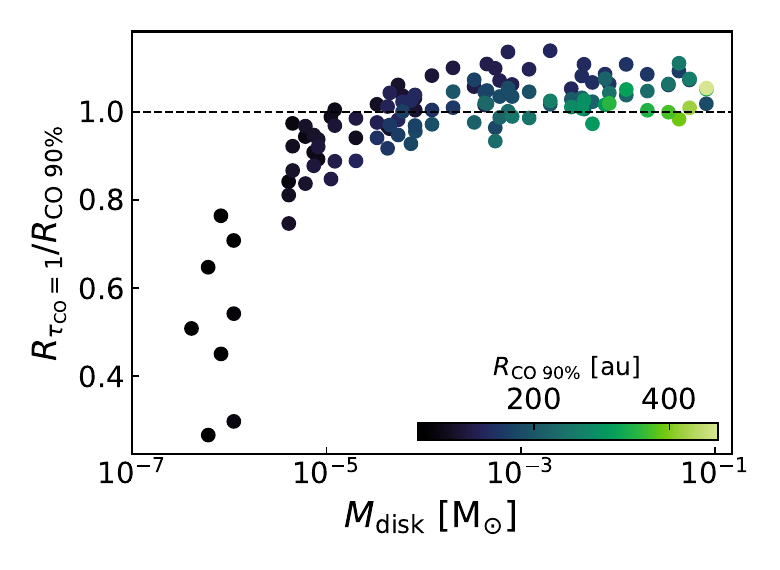}
    \end{minipage}    
    \caption{\label{appfig: Ngas at Rtau and ratio of Rtau and Rco} \textbf{Left:} comparison between $N_{\rm gas}(R_{\tau})$ as derived from Eq. (A.12) and the value of $N_{\rm gas}(R_{\tau})$ as obtained from the \texttt{DALI} models. For the analytical $N_{\rm gas}$, $x_{\rm CO,peak} = 3\times10^{-5}, N_{\rm CO} = 3\times10^{15}\ \mathrm{cm}^{-2}$ and $z_{\rm freeze}(R_{\tau})$ was also obtained from the models. Colors show the value of $R_{\tau}$.
    \textbf{Right:} Ratio of $R_{\tau}$ and \rgas\ set against the disk mass.}
\end{figure*}

As a proof-of-concept we obtain $z_{\rm freeze}(R_{\tau})$ from our models, in favor of the increased complexity that a full fit of the temperature structure would bring, and combine it with informed values of $x_{\rm CO,peak}=3\times10^{-5}$ and $N_{\rm CO} = 3\times10^{15}\ \mathrm{cm}^{-2}$ to calculate $N_{\rm gas}(R_{\tau})$ using Eq. \eqref{eq: numerical equation}. The left panel of Figure \ref{appfig: Ngas at Rtau and ratio of Rtau and Rco} shows that these analytical $N_{\rm gas}$ reproduce the values from the models, including its dependence on disk mass. However, the figure also shows that the relation between $N_{\rm gas}(R_{\tau})$ and \mdisk\ is not a powerlaw as it is for $N_{\rm gas}(\rgas)$ and it is also less tight. The underlying cause for this is the fact that the relation between \rgas\ and $R_{\tau}$ also depends on disk mass. The right panel of Figure \ref{appfig: Ngas at Rtau and ratio of Rtau and Rco} shows the ratio $R_{\tau}/\rgas$, which decreases towards lower disk mass. This mass dependence might appear small, but one should bear in mind that surface density at these radii follows an exponential; a small difference in radius will correspond to a much larger difference in gas column density. This effect introduces a further mass dependence, as more massive disks have a larger \rgas\ (and $R_{\tau}$) that lies further in the exponential taper of the surface density profile where it is steeper, meaning that differences between \rgas\ and $R_{\tau}$ will result in larger differences between $N_{\rm gas}(\rgas)$ and $N_{\rm gas}(R_{\tau})$ for more massive disks. This is a complex process to model, prompting us to use the empirical correlation presented in Section \ref{sec: results}.

\section{Effect of disk and stellar parameters on the height of the CO snow surface}
\label{app: zfreeze-mdisk correlation}

\begin{figure*}
    \centering
    \includegraphics[width=\textwidth]{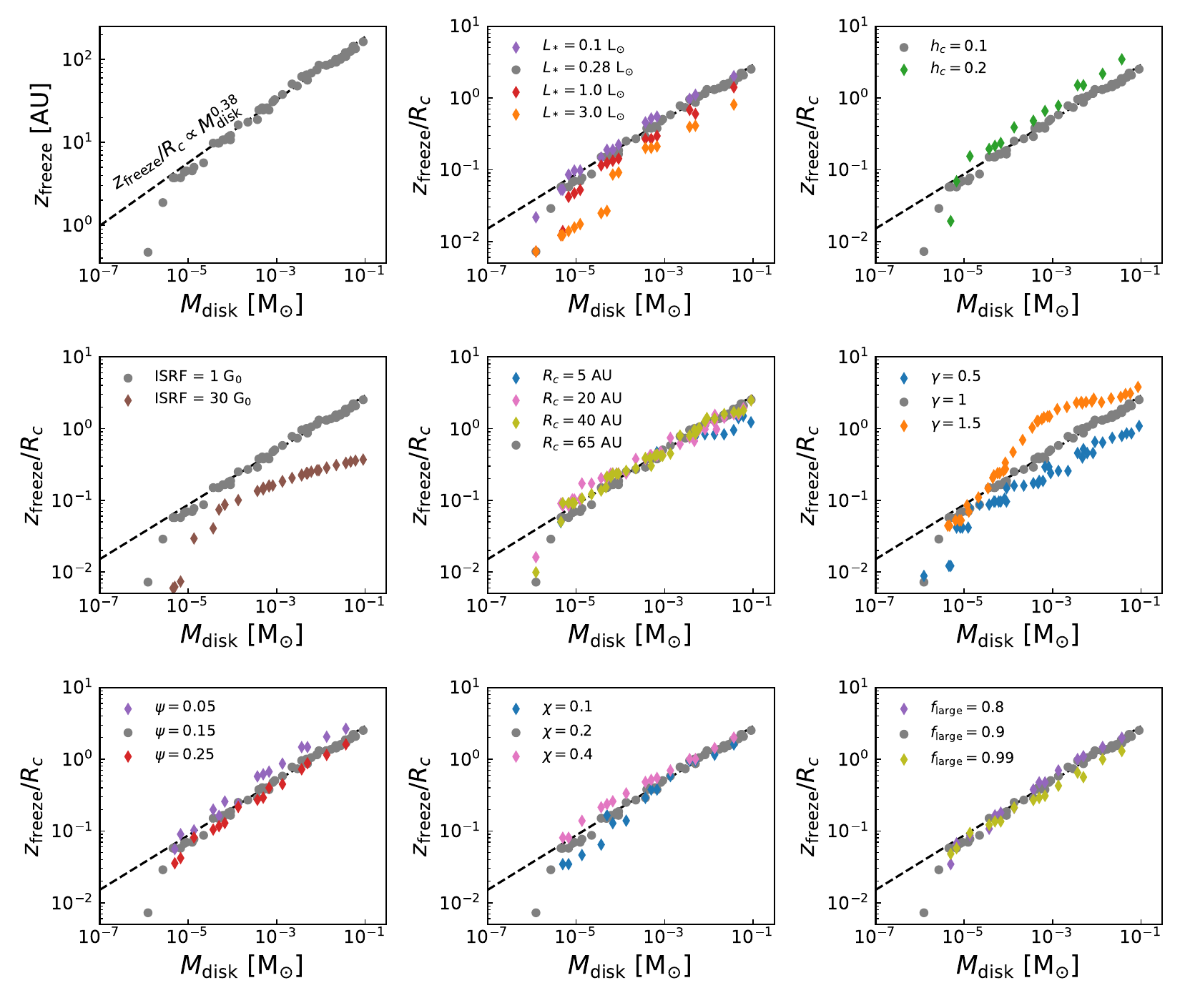}
    \caption{\label{appfig: zfreeze-mdisk correlation} Height of the CO snow-surface $(z_{\rm freeze})$ at \rgas\ versus disk mass. From left to right, top to bottom we show the effect of stellar luminosity$(L_*)$, the scale height at \rc\ $(h_c)$, the external interstellar radiation field (ISRF), the characteristic size $(\rc)$, the slope of the surface density $(\gamma)$, the disk flaring angle $(\psi)$, the scale height reduction of the large grains $(\chi)$ and the fraction of large grains $(f_{\rm large})$. The gray points in each panel show the fiducial models shown in Figure \ref{fig: Ngas-mdisk correlation}. The black dashed line shows $z_{\rm freeze}(\rgas)/\rc\propto\mdisk^{0.38}$.}
\end{figure*}

In Figure \ref{appfig: zfreeze-mdisk correlation} from left to right, top to bottom the examined parameters are the stellar luminosity$(L_*)$, the scale height at \rc\ $(h_c)$, the external interstellar radiation field (ISRF), the characteristic size $(\rc)$, the slope of the surface density $(\gamma)$, the disk flaring angle $(\psi)$, the scale height reduction of the large grains $(\chi)$ and the fraction of large grains $(f_{\rm large})$. The gray points in each panel show the fiducial models shown in Figure \ref{fig: Ngas-mdisk correlation}. The black dashed line shows $N_{\rm gas}(\rgas)\propto\mdisk^{0.34}$.

\section{Measuring the observed disk radius using 68\% instead of 90\% of the CO flux}
\label{app: R68}

\begin{figure*}[htb]
    \centering
    \includegraphics[width=\textwidth]{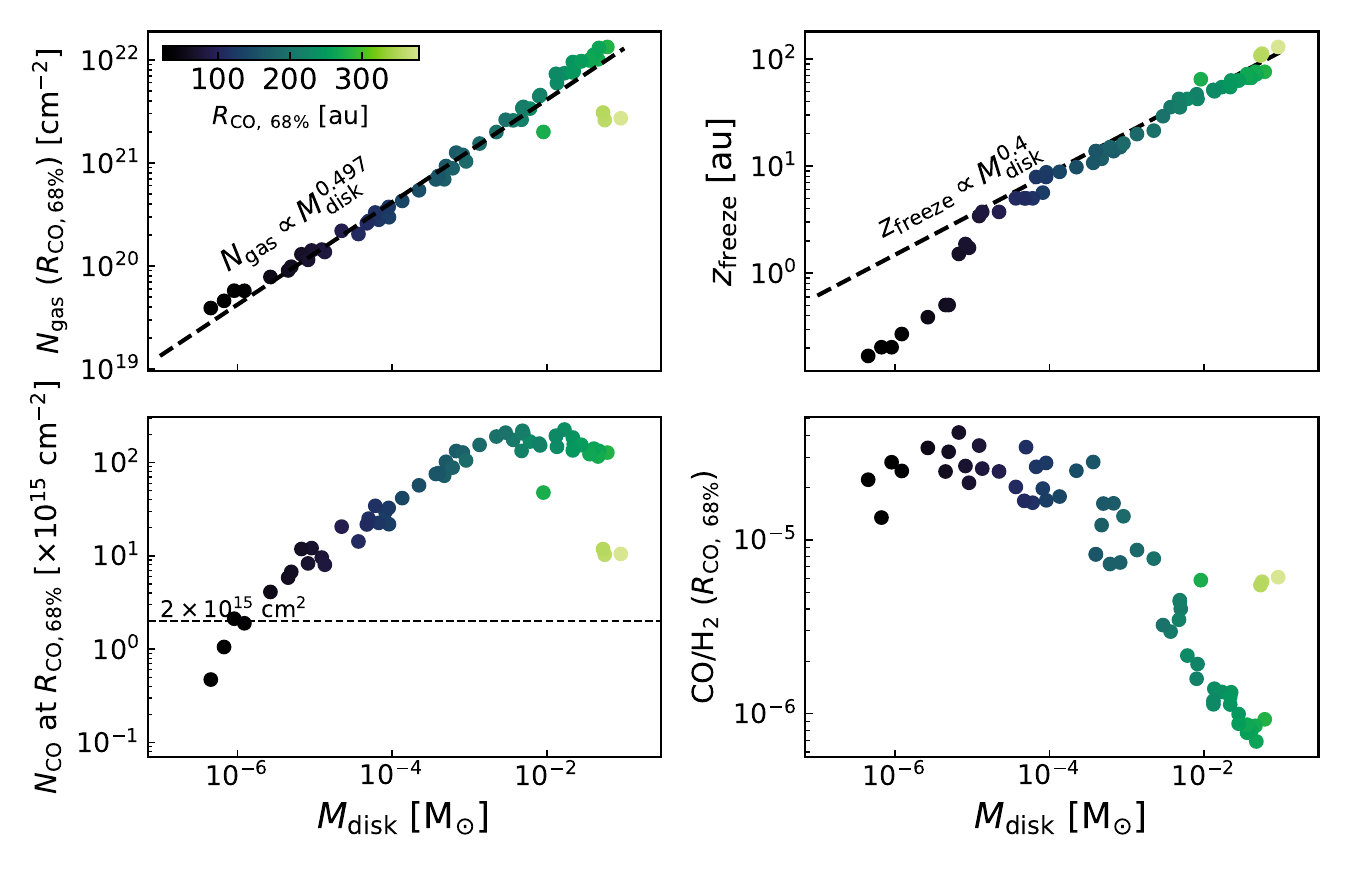}
    \caption{\label{fig: Ngas-mdisk correlation -- R68} \newtext{As Figure \ref{fig: Ngas-mdisk correlation}, but each property is now measured at the radius that encloses 68\% of the CO emission.} 
    }
\end{figure*}

\newtext{Throughout this work we have used \rgas\ as an observational measure of the disk size, but tests show that a similar result, at least qualitatively, can also be obtained if we instead use the radius that encloses 68\% of the CO 2-1 flux $(R_{\rm CO,\ 68\%})$. Recreating Figure \ref{fig: Ngas-mdisk correlation} but now for disk properties measured at $R_{\rm CO,\ 68\%}$, we find that there exists a similar powerlaw relation between $N_{\rm gas}(R_{\rm CO,\ 68\%})$ and \mdisk\ as there did for \rgas. While there is much less of a direct link between $R_{\rm CO,\ 68\%}$ and the radius where CO becomes photo-dissociated, a fact that can be gleaned from the wide range of $N_{\rm CO}$ at $R_{\rm CO,\ 68\%}$, we find a tight relation between \rgas\ and $R_{\rm CO,\ 68\%}$ in our models which allows us to also relate $R_{\rm CO,\ 68\%}$ to the radius where CO becomes photo-dissociated. The tight relation between \rgas\ and $R_{\rm CO,\ 68\%}$ reflects to overall similarity in CO emission profiles between our models, suggesting that our findings for \rgas\ and $R_{\rm CO,\ 68\%}$ likely hold for most fraction-of-CO-flux-radii.}

\newtext{If we fit a powerlaw to $N_{\rm gas}(R_{\rm CO,\ 68\%})$ and the disk mass we obtain the following critical gas column density}
\begin{equation}
\label{eq: gas column disk mass relation -- 68}
N_{\rm gas}(R_{\rm CO,\ 68\%}) \approx 4.1\times10^{22} \left(\frac{M_{\rm gas}}{\mathrm{M}_{\odot}}\right)^{0.5}\ \mathrm{cm}^{-2}.   
\end{equation}

\newtext{Using this critical column density instead of the one for \rgas\ changes the square bracket term in Equation \eqref{eq: full prescription} to}
\begin{align}
    [..] &= \frac{\Sigma_{\rm c}}{\mu m_H N_{\rm gas}(R_{\rm CO,\ 68\%})}\\
         &= 4.5\times10^6 \left(\frac{\mdisk}{\msun}\right)^{0.5}\left(\frac{\rc}{\rm au}\right)^{-2},
\end{align}

\newtext{which gives us the following analytical prescription for $R_{\rm CO,\ 68\%}$}
\begin{equation}
    \label{eq: full prescription 68}
        R_{\rm CO,\ 68\%} = R_c \left(\frac{\gamma-\xi}{2-\gamma} W\left(\frac{2 -\gamma}{\gamma-\xi} \left[ 4.5\cdot10^6 \left(\frac{M_{\rm d}}{\mathrm{M}_{\odot}}\right)^{0.5} \left(\frac{R_c}{\rm au}\right)^{-2} \right]
        ^{\frac{2-\gamma}{\gamma -\xi}}\right)  \right)^{\frac{1}{2-\gamma}}
\end{equation}.

\section{Deriving a minimum disk mass based on \rgas}
\label{app: deriving minimum mass}

In Section \ref{sec: rgas versus rc} we showed that there is a minimum disk mass associated with each \rgas. Here we derive this mass analytically.
We begin with the analytical formula for \rgas\ (Eq. \eqref{eq: full prescription})

\begin{align}
\label{eq: minimum mass derivation - step 1}
        R_{\rm CO,\ 90\%} &= R_c \left(\frac{\gamma-\xi}{2-\gamma} W\left(\frac{2 -\gamma}{\gamma-\xi} \left[ 4.9\cdot10^7  \left(\frac{M_{\rm d}}{\mathrm{M}_{\odot}}\right)^{0.66} \left(\frac{\rm au}{R_c}\right)^{2} \right]
        ^{\frac{2-\gamma}{\gamma -\xi}}\right)  \right)^{\frac{1}{2-\gamma}}\\
                          &= \rc \left[\frac{1}{a}W(x)\right]^{1/(2-\gamma)}.
\end{align}
Here we have defined a few short hands $x = a y^a$, $y = 4.9\cdot10^7  \left(\frac{M_{\rm d}}{\mathrm{M}_{\odot}}\right)^{0.66} \left(\frac{R_c}{\rm au}\right)^{-2}$ and $a = \tfrac{2-\gamma}{\gamma-\xi}$. Taking the derivative of \rgas\ to \rc\  and setting it to zero we obtain

\begin{align}
\label{eq: minimum mass derivation - step 2}
\frac{\partial\ \rgas}{\partial \rc} &= \left[\frac{1}{a}W(x)\right]^{1/(2-\gamma)} + \rc \frac{\partial}{\partial \rc}\left[\frac{1}{a}W(x)\right]^{1/(2-\gamma)}\\
 &= \left[\frac{1}{a}W(x)\right]^{\frac{1}{2-\gamma}} + \frac{\rc}{2-\gamma} \left[\frac{1}{a}W(x)\right]^{\frac{\gamma -1}{2-\gamma}} \frac{\partial\ W(x)}{\partial x}\frac{\partial\ x}{\partial y}\frac{\partial\ y}{\partial\rc}\\
 &= \left[\frac{1}{a}W(x)\right]^{\frac{1}{2-\gamma}} - 2\frac{ax}{2-\gamma} \left[\frac{1}{a}W(x)\right]^{\frac{\gamma-1}{2-\gamma}} \frac{\partial\ W(x)}{\partial x} \\
 &= \left[\frac{1}{a}W(x)\right]^{\frac{1}{2-\gamma}} - 2\frac{ax}{2-\gamma} \left[\frac{1}{a}W(x)\right]^{\frac{\gamma-1}{2-\gamma}} \frac{W(x)}{x\left(1+W(x)\right)} \\
0 &= 1 - \frac{2a}{2-\gamma} \left[\frac{1}{a}W(x)\right]^{-1} \frac{W(x)}{\left(1+W(x)\right)}\\
\frac{2-\gamma}{2a^2} &=  \frac{1}{\left(1+W(x)\right)}\\
W(x) &= \frac{2a^2}{2-\gamma}-1
\end{align}
The inverse Lambert-W function is given by $W^{-1}(y) = ye^y$. With this and our shorthands we can write out the maximum \rgas, and its corresponding \rc, for a given mass $M_{\rm disk}$
\begin{align}
\label{eq: minimum mass derivation -- sizes}
&R_{\rm c,\ max} = \sqrt{4.9\cdot10^7 M_{\rm disk}^{0.66}} \left[\frac{1}{a} W^{-1}\left(\frac{2a^2}{2-\gamma}-1\right)\right]^{-\frac{1}{2a}}\\
&R_{\rm CO,\ 90\%,max} = R_{\rm c,\ max} \left[\frac{1}{a}\left(\frac{2a^2}{2-\gamma}-1\right)\right]^{1/(2-\gamma)}
\end{align}
For $\gamma=1,\xi=0$ these equations reduce to
\begin{equation}
\label{eq: minimum mass derivation -- sizes simple}
R_{\rm CO,\ 90\%,max} = R_{\rm c,\ max} = \sqrt{\frac{4.9\cdot10^7 M_{\rm disk}^{0.66}}{e}}\\
\end{equation}
Writing the disk mass in terms of $R_{\rm CO,\ 90\%,max}$ then gives Eq. \eqref{eq: mass lower limit}
\begin{equation}
\label{eq: minimum mass derivation -- result}
\mdisk \geq \left[\frac{e R_{\rm CO,\ 90\%,max}}{4.9\cdot10^7}\right]^{1/0.66} \gtrsim 1.295\times10^{-5}\left(\frac{\rgas}{\rm 100\ au}\right)^3\ \msun.
\end{equation}

\section{Deriving \rc\ for disks in Lupus, Upper Sco, Taurus and DSHARP}
\label{app: derived rc}

\begin{figure}[htb]
    \centering
    \includegraphics[width=\columnwidth]{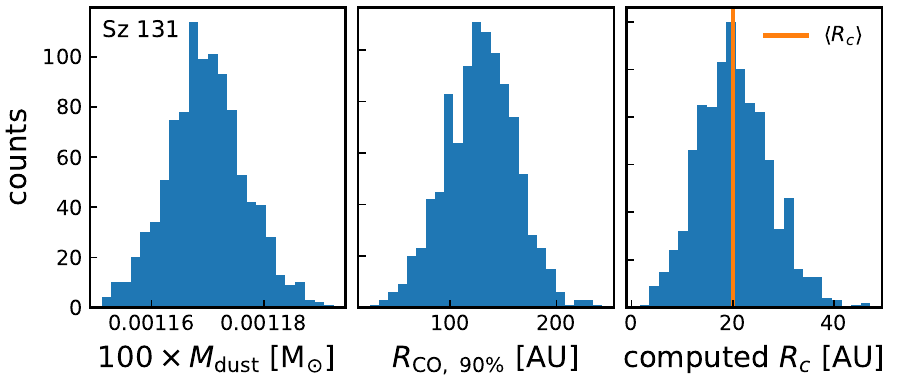}
    \caption{\label{fig: error prop} Example of how \rc\ is derived. A total of 1000 samples is drawn from the dust mass $M_{\rm dust}$ and observed gas disk size \rgas\ of Sz 131 based on the Gaussian uncertainties found for these two properties (leftmost and center panel). These samples ($100\times M_{\rm dust},\rgas$) are then used to calculate the corresponding $R_{c,i}$, resulting the distribution of \rc\ shown in the rightmost panel. }
\end{figure}

\begin{figure}[htb]
    \centering
    \includegraphics[width=\columnwidth]{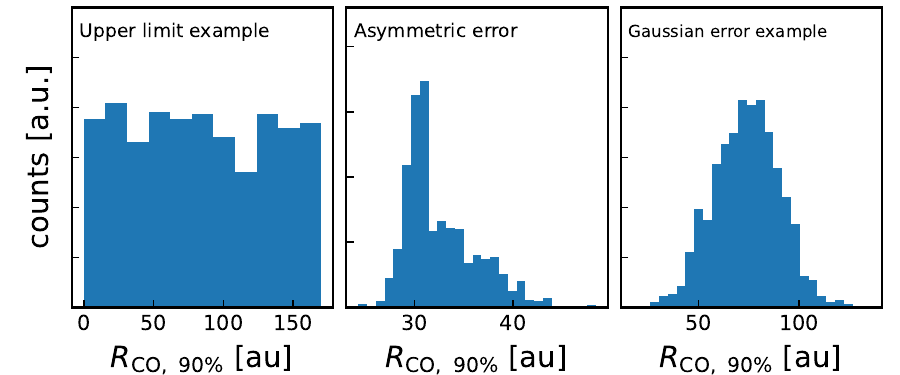}
    \caption{\label{fig: Rco error examples} Examples of the three types of uncertainties on \rgas\ in our sample. \textbf{Left}: upper limits on \rgas\ are represented by a uniform distribution between 0 and the upper limit. \textbf{Middle}: The asymmetrical uncertainties on \rgas\ in Upper Sco are represented by two half-Gaussians with different width. \textbf{Right}: for the majority of sources the uncertainties on \rgas\ follow a Gaussian distribution.}
\end{figure}

Our approach for deriving \rc\ is as follows (shown in Figure \ref{fig: error prop}). We collected a sample of disks with a measured \rgas\ or an upper limit on \rgas\ from the literature (\citealt{barenfeld2017,ansdell2018,Sanchis2021,Long2022}, see Table \ref{tab: observed sample}). For each source in this sample we first draw a random $\mdisk=100\times\mdust$ from the distribution of the observed \mdust\ and its uncertainties. We do the same for \rgas, where upper limits on \rgas\ are treated as a uniform distribution between 0 and the upper limit. For Upper Sco \rgas\ is calculated from fitted CO intensity profile reported in Table 4 in \cite{barenfeld2017}. Note that the uncertainties on the intensity profile are asymmetrical, which when propagated into the uncertainty on \rgas\ is represented by a two half-Gaussians with different width (see Figure \ref{fig: Rco error examples}).  
For this $(\mdisk,\rgas)_i$, we calculate \rc\ by inverting Equation \eqref{eq: full prescription}, where we assume that $\rgas \gg \rc$ (see Section \ref{sec: rgas versus rc}). This procedure is repeated $N=1000$ times to properly sample the distribution of \rc\ of each source. Table \ref{tab: observed sample} lists the derived \rc\ and its uncertainties for each source.

\begin{table*}[htb]
\centering
\caption{\label{tab: observed sample} Derived \rc\ for Lupus, Upper Sco, Taurus and DSHARP}
\def\arraystretch{1.1}
\begin{tabular*}{0.65\textwidth}{lc|cc|cc}
\hline\hline
Name  &  sample  &  $M_{\rm dust}$ & \rgas\ & \rc\ & ref\\
      &          &  [M$_{\oplus}$]  &  [au]  & [au] &\\
\hline
EX Lup & Lupus &19.1$\pm$0.4 & $\leq$170.3 &$3.8^{+6.0}_{-5.4}$ &(2,3,4)\\
 Lup706 & Lupus &0.4$\pm$0.0 & $\leq$87.2 &$2.6^{+6.1}_{-4.0}$ &(1,3,4)\\
RXJ1556.1-3655 & Lupus &24.8$\pm$0.1 &118.5$\pm$11.2 &$13.1^{+1.1}_{-1.2}$ &(2,3,4)\\
RY Lup & Lupus &123.0$\pm$0.3 &323.0$\pm$96.9 &$29.7^{+12.5}_{-10.2}$ &(1,3,4)\\
Sz 65 & Lupus &27.4$\pm$0.1 &167.7$\pm$19.9 &$18.5^{+2.3}_{-2.3}$ &(1,3,4)\\
Sz 66 & Lupus &6.5$\pm$0.1 & $\leq$135.3 &$3.4^{+5.4}_{-4.6}$ &(1,3,4)\\
Sz 69 & Lupus &7.1$\pm$0.1 &123.7$\pm$17.6 &$15.9^{+2.7}_{-2.2}$ &(1,3,4)\\
Sz 72 & Lupus &6.0$\pm$0.1 &32.7$\pm$11.1 &$2.3^{+0.9}_{-0.8}$ &(1,3,4)\\
Sz 73 & Lupus &13.2$\pm$0.1 &106.6$\pm$13.3 &$11.5^{+1.4}_{-1.3}$ &(1,3,4)\\
Sz 75 & Lupus &31.9$\pm$0.1 &226.2$\pm$67.9 &$22.1^{+7.1}_{-7.8}$ &(1,3,4)\\
Sz 76 & Lupus &4.9$\pm$0.2 &140.4$\pm$13.6 &$19.7^{+2.1}_{-1.9}$ &(2,3,4)\\
Sz 77 & Lupus &2.1$\pm$0.1 &37.2$\pm$17.6 &$2.4^{+1.6}_{-1.6}$ &(2,3,4)\\
Sz 83 & Lupus &191.7$\pm$0.2 & $\leq$347.9 &$7.9^{+11.5}_{-10.3}$ &(1,3,4)\\
Sz 84 & Lupus &13.4$\pm$0.1 &192.3$\pm$25.9 &$26.8^{+3.9}_{-3.4}$ &(1,3,4)\\
Sz 90 & Lupus &9.9$\pm$0.2 &75.4$\pm$22.8 &$6.5^{+2.2}_{-2.1}$ &(1,3,4)\\
Sz 91 & Lupus &27.7$\pm$0.5 &330.9$\pm$99.3 &$40.4^{+17.7}_{-16.4}$ &(1,3,4)\\
Sz 96 & Lupus &1.8$\pm$0.1 &32.9$\pm$15.5 &$2.2^{+1.3}_{-1.3}$ &(1,3,4)\\
Sz 100 & Lupus &18.1$\pm$0.2 &128.7$\pm$23.3 &$14.6^{+2.9}_{-2.6}$ &(1,3,4)\\
Sz 102 & Lupus &6.1$\pm$0.4 &74.5$\pm$45.0 &$6.2^{+6.2}_{-4.9}$ &(2,3,4)\\
Sz 111 & Lupus &79.3$\pm$0.4 &459.1$\pm$137.7 &$56.2^{+24.6}_{-22.8}$ &(1,3,4)\\
Sz 114 & Lupus &44.8$\pm$0.2 &170.3$\pm$34.5 &$18.2^{+3.7}_{-3.6}$ &(1,3,4)\\
Sz 118 & Lupus &30.0$\pm$0.4 &145.9$\pm$32.6 &$14.4^{+3.7}_{-2.8}$ &(1,3,4)\\
Sz 130 & Lupus &2.8$\pm$0.1 &120.2$\pm$27.3 &$15.3^{+4.4}_{-3.5}$ &(1,3,4)\\
Sz 131 & Lupus &3.9$\pm$0.1 &128.2$\pm$34.1 &$15.3^{+5.1}_{-4.7}$ &(1,3,4)\\
Sz 133 & Lupus &28.5$\pm$0.2 &206.7$\pm$23.9 &$28.1^{+3.8}_{-3.8}$ &(1,3,4)\\
J154518.5-342125 & Lupus &2.3$\pm$0.2 &36.4$\pm$12.9 &$3.0^{+1.4}_{-1.2}$ &(1,3,4)\\
J160002.4-422216 & Lupus &57.0$\pm$0.1 &261.1$\pm$30.4 &$34.0^{+3.8}_{-4.5}$ &(1,3,4)\\
J160703.9-391112 & Lupus &2.0$\pm$0.2 &225.1$\pm$67.5 &$51.6^{+57.4}_{-30.1}$ &(1,3,4)\\
J160830.7-382827 & Lupus &58.2$\pm$0.5 &343.4$\pm$103.0 &$34.6^{+14.2}_{-12.3}$ &(1,3,4)\\
J160901.4-392512 & Lupus &8.3$\pm$0.3 &193.9$\pm$18.7 &$30.3^{+3.0}_{-3.2}$ &(1,3,4)\\
J160927.0-383628 & Lupus &1.7$\pm$0.1 &113.1$\pm$20.4 &$16.1^{+3.2}_{-3.1}$ &(1,3,4)\\
J161029.6-392215 & Lupus &3.4$\pm$0.1 &133.8$\pm$25.5 &$18.0^{+4.3}_{-3.8}$ &(1,3,4)\\
J161243.8-381503 & Lupus &13.5$\pm$0.2 &67.1$\pm$24.9 &$4.9^{+2.3}_{-2.3}$ &(1,3,4)\\
V1094Sco & Lupus &230.3$\pm$8.4 &420.9$\pm$32.7 &$50.6^{+3.5}_{-3.9}$ &(2,3,4)\\
V1192Sco & Lupus &0.4$\pm$0.1 & $\leq$226.2 &$8.1^{+21.6}_{-15.0}$ &(1,3,4)\\
J16070854-3914075 & Lupus &50.2$\pm$0.6 &339.3$\pm$47.5 &$45.4^{+8.1}_{-6.9}$ &(1,3,4)\\
\hline\hline
.... & & & & &\\
\hline
\end{tabular*}
\begin{minipage}{0.65\textwidth}
\vspace{0.1cm}
{\footnotesize{$^1$\citealt{ansdell2018},$^2$\citealt{ansdell2016},$^3$\citealt{Sanchis2021},$^4$\citealt{Alcala2019},$^5$\citealt{Barenfeld2016},$^6$\citealt{barenfeld2017},$^7$\citealt{Facchini2019},$^8$\citealt{Long2018},$^9$\citealt{Flaherty2020},$^{10}$\citealt{Pegues2021},$^{11}$\citealt{Kurtovic2021},$^{12}$\citealt{Long2022},$^{13}$\citealt{Andrews2018}}}
\end{minipage}
\end{table*}

\begin{table*}[htb]
\centering
\caption{\label{tab: observed sample 2} Derived \rc\ for Lupus, Upper Sco, Taurus and DSHARP, cont'd}
\def\arraystretch{1.1}
\begin{tabular*}{0.65\textwidth}{lc|cc|cc}
\hline\hline
Name  &  sample  &  $M_{\rm dust}$ & \rgas\ & \rc\ & ref\\
      &          &  [M$_{\oplus}$]  &  [au]  & [au] &\\
\hline
.... & & & & &\\
J16081497-3857145 & Lupus &3.7$\pm$0.1 &95.1$\pm$31.5 &$10.9^{+5.5}_{-4.9}$ &(1,3,4)\\
J16085953-3856275 & Lupus &0.2$\pm$0.0 & $\leq$36.0 &$0.8^{+1.5}_{-1.3}$ &(1,3,4)\\
\hline\hline
\hline\hline
CXTau & Taurus &4.8$\pm$0.5 &115.0$\pm$13.0 &$14.8^{+1.4}_{-1.7}$ &(7,12)\\
DLTau & Taurus &130.1$\pm$13.0 &597.0$\pm$91.0 &$82.8^{+14.9}_{-13.3}$ &(8,12)\\
DMTau & Taurus &49.9$\pm$5.0 &876.0$\pm$23.0 &$246.6^{+9.2}_{-13.2}$ &(12)\\
GOTau & Taurus &34.2$\pm$3.4 &1014.0$\pm$83.0 &$325.5^{+54.7}_{-42.0}$ &(8,12)\\
UZTau & Taurus &67.0$\pm$6.7 &389.0$\pm$75.0 &$47.2^{+10.7}_{-9.8}$ &(8,12)\\
FPTau & Taurus &4.1$\pm$0.4 &74.0$\pm$17.0 &$7.8^{+1.8}_{-1.7}$ &(10,12)\\
CIDA1 & Taurus &13.3$\pm$1.3 &132.0$\pm$14.0 &$16.2^{+1.8}_{-1.7}$ &(11,12)\\
CIDA7 & Taurus &9.5$\pm$0.9 &95.0$\pm$11.0 &$11.1^{+1.3}_{-1.3}$ &(11,12)\\
MHO6 & Taurus &19.8$\pm$2.0 &218.0$\pm$7.0 &$35.2^{+1.2}_{-0.9}$ &(11,12)\\
J0415 & Taurus &0.4$\pm$0.0 &47.0$\pm$13.0 &$5.1^{+1.6}_{-1.8}$ &(11,12)\\
J0420 & Taurus &9.6$\pm$1.0 &59.0$\pm$10.0 &$5.8^{+1.0}_{-0.9}$ &(11,12)\\
J0433 & Taurus &22.5$\pm$2.3 &165.0$\pm$12.0 &$21.7^{+1.5}_{-1.7}$ &(11,12)\\
\hline\hline
GW Lup & DSHARP &60.5$\pm$6.1 &267.0$\pm$8.0 &$36.2^{+1.1}_{-1.0}$ &(13,12)\\
IM Lup & DSHARP &178.8$\pm$17.9 &803.0$\pm$9.0 &$123.2^{+0.0}_{-2.2}$ &(13,12)\\
MY Lup & DSHARP &54.4$\pm$5.4 &192.0$\pm$7.0 &$18.3^{+1.9}_{-2.3}$ &(13,12)\\
Sz 129 & DSHARP &63.1$\pm$6.3 &130.0$\pm$8.0 &$16.5^{+3.5}_{-3.5}$ &(13,12)\\
AS209 & DSHARP &119.4$\pm$11.9 &280.0$\pm$5.0 &$33.1^{+0.6}_{-0.6}$ &(13,12)\\
SR4 & DSHARP &35.1$\pm$3.5 &82.0$\pm$7.0 &$7.5^{+0.6}_{-0.6}$ &(13,12)\\
DoAr25 & DSHARP &132.7$\pm$13.3 &233.0$\pm$6.0 &$26.4^{+0.7}_{-0.5}$ &(13,12)\\
DoAr33 & DSHARP &19.1$\pm$1.9 &64.0$\pm$6.0 &$5.7^{+0.5}_{-0.5}$ &(13,12)\\
WaOph6 & DSHARP &69.0$\pm$6.9 &297.0$\pm$7.0 &$36.2^{+0.7}_{-0.9}$ &(13,12)\\
HD142666 & DSHARP &74.4$\pm$7.4 &171.0$\pm$5.0 &$16.8^{+0.4}_{-0.4}$ &(13,12)\\
HD143006 & DSHARP &45.5$\pm$4.5 &154.0$\pm$5.0 &$16.0^{+0.5}_{-0.4}$ &(13,12)\\
HD163296 & DSHARP &206.5$\pm$20.7 &478.0$\pm$5.0 &$54.1^{+0.3}_{-0.6}$ &(13,12)\\
\hline
\end{tabular*}
\begin{minipage}{0.65\textwidth}
\vspace{0.1cm}
{\footnotesize{$^1$\citealt{ansdell2018},$^2$\citealt{ansdell2016},$^3$\citealt{Sanchis2021},$^4$\citealt{Alcala2019},$^5$\citealt{Barenfeld2016},$^6$\citealt{barenfeld2017},$^7$\citealt{Facchini2019},$^8$\citealt{Long2018},$^9$\citealt{Flaherty2020},$^{10}$\citealt{Pegues2021},$^{11}$\citealt{Kurtovic2021},$^{12}$\citealt{Long2022},$^{13}$\citealt{Andrews2018}}}
\end{minipage}
\end{table*}

\end{appendix}
\end{document}